%% file: ms.tex
\documentclass[twocolumn]{aastex631}
\usepackage{amsmath,amssymb,epsfig,epsfig,xspace,graphicx,microtype}

\usepackage{orcidlink}

\usepackage[caption=false]{subfig}

\usepackage{natbib}
\usepackage{float}
\usepackage{xcolor}

\begin{document}

\newcommand{\degrees}{$^{\circ}$}

\def\mpch {$h^{-1}$ Mpc} 
\def\kpch {$h^{-1}$ kpc} 
\def\kms {km s$^{-1}$} 

\def\lcdm {$\Lambda$CDM } 
\newcommand{\hMpc}{\ensuremath{\,h^{-1}\, \textrm{Mpc}}}
\def\etal {et al.}
\def\kt{\tilde{k}}
\def\mpc{{\rm Mpc}}
\def\msunyr{\hbox{$M_{\odot}~{\rm yr}^{-1}$}}
\def\deg{\hbox{$^{\circ}$}}
\newcommand{\mass}{\ensuremath{\mathrm{M_*}}}
\newcommand{\mstar}{\ensuremath{\mathcal{M}_*}}
\newcommand{\msun}{\ensuremath{M_{\odot}}}
\newcommand{\logmass}{\ensuremath{\log\,(\mass/\msun)}}
\newcommand{\sfr}{\ensuremath{\psi}}
\newcommand{\sfrm}{{\rm SFR}/\ensuremath{\mass}}
\newcommand{\sfrunit}{\ensuremath{\mathcal{M}_{\sun}~\textrm{yr}^{-1}}}
\newcommand{\arcsecsq}{\ensuremath{\mathrm{arcsec}^2}\xspace}
\newcommand{\vig}{\sigma}

\newcommand{\um}{\ensuremath{\mu\mathrm{m}}\xspace}
\newcommand{\uJy}{\ensuremath{\mu\mathrm{Jy}}\xspace}
\newcommand{\degsq}{\ensuremath{\mathrm{deg}^2}}
\newcommand{\ssfr}{\ensuremath{\mathrm{sSFR}}\xspace}
\newcommand{\ssfrunit}{\ensuremath{\mathrm{yr}^{-1}}}
\newcommand{\oi}{[\textrm{O}~\textsc{i}]}
\newcommand{\oii}{[\textrm{O}~\textsc{ii}]}
\newcommand{\oiii}{[\textrm{O}~\textsc{iii}]}
\newcommand{\sii}{[\textrm{S}~\textsc{ii}]}
\newcommand{\nii}{[\textrm{N}~\textsc{ii}]}
\newcommand{\hei}{[\textrm{He}~\textsc{i}]}
\newcommand{\heii}{[\textrm{He}~\textsc{ii}]}
\newcommand{\mgii}{\textrm{Mg}~\textsc{ii}}
\newcommand{\mgi}{\textrm{Mg}~\textsc{i}}
\newcommand{\neiii}{[\textrm{Ne}~\textsc{iii}]}
\newcommand{\nev}{[\textrm{Ne}~\textsc{v}]}
\newcommand{\hb}{H$\beta$}
\newcommand{\hg}{H$\gamma$}
\renewcommand{\kms}{\ensuremath{\mathrm{km\,s^{-1}}}}

\newcommand{\vcen}{$v_{50}$}
\newcommand{\vsig}{$\sigma$}
\newcommand{\vmax}{$v_{max}$}

\title{KCWI Observations of Type 2 Quasar-driven Ionized Outflows at $z\sim0.1$}

\correspondingauthor{Alison Coil}
\email{acoil@ucsd.edu}

\author[0000-0002-2583-5894]{Alison L. Coil}
\affiliation{Department of Astronomy and Astrophysics, University of California, 9500 Gilman Dr., La Jolla, CA 92093}

\author[0009-0007-6748-4627]{Saloni Agrawal}
\affiliation{Department of Astronomy and Astrophysics, University of California, 9500 Gilman Dr., La Jolla, CA 92093}

\author[0000-0002-1608-7564]{David S.~N. Rupke}
\affiliation{Department of Physics, Rhodes College, 2000 N. Parkway, Memphis, TN 38112}

\author[0000-0002-2451-9160]{Serena Perrotta}
\affiliation{Department of Astronomy and Astrophysics, University of California, 9500 Gilman Dr., La Jolla, CA 92093}

\author[0000-0001-9489-3791]{Natalie Lam}
\affiliation{Department of Physics and Astronomy, University of California, Los Angeles, 430 Portola Plaza, Los Angeles, CA 90095}

\author[0000-0002-6121-3443]{Wenmeng Ning}
\affiliation{Department of Physics and Astronomy, University of California, Los Angeles, 430 Portola Plaza, Los Angeles, CA 90095}

\author[0009-0006-4065-713X]{Casey Carlile}
\affiliation{Department of Astronomy and Astrophysics, University of California, 9500 Gilman Dr., La Jolla, CA 92093}

\author[0000-0003-4209-1599]{Yu-Hsuan Teng}
\affiliation{Department of Astronomy and Joint Space-Science Institute, University of Maryland, 4296 Stadium Drive, College Park, MD 20742}


\begin{abstract}

AGN-driven outflows play a crucial role in regulating star formation in galaxies. These outflows are known to be multi-phase, and the characterization of the kinematics, extent, and energetics of each gas phase is needed to obtain a full picture of the outflow properties and impact. This study presents Keck Cosmic Web Imager (KCWI) integral field spectroscopy of the ionized gas outflow properties in seven low redshift ($z<0.1$) quasars---six Type 2 AGN and one narrow-line Type 1 AGN---providing insight into the warm ionized outflows in these luminous AGN. With the large field-of-view and high sensitivity of KCWI, these data reveal both larger ionized outflow extents than were previously known for these sources and uncover the existence of additional high-velocity kinematic components in the ionized outflows of these luminous AGN.  We find (a) a high incidence (50-70\%) of multiple outflows on scales of 4--10 kpc and (b) trends between the outflow extent, kinematics, and the ionized mass and energy outflow rates. These trends imply that the faster and often smaller outflows carry higher mass and energy outflow rates than the slower, larger outflows.  We also find that the ionized gas outflows reach as far as the known extent of narrow-line ionized gas emission in AGN, which is known to depend weakly on AGN luminosity.  This work highlights the utility of the current generation of wide-field integral field spectrographs on 8--10m class telescopes for studying quasar-driven outflows.

\end{abstract}

\section{Introduction}

AGN feedback is an essential element of present-day models of galaxy formation and evolution \citep[e.g.,][]{DiMatteo05, Booth09, Dave11, Genel14, Schaye15, Richardson16, Beckmann17, Torrey20}. AGN feedback 
in the form of wide-angle winds regulates and may quench star formation in galaxies \citep{Zubovas12, Pontzen17}, shape galaxy morphology \citep{Dubois16}, impact the distribution of the  Ly$\alpha$ forest
\citep{Gurvich17}, and regulate supermassive black hole (SMBH)
accretion \citep{Hopkins16, Volonteri16}. 

AGN-driven outflows are common across a wide range of redshifts. Large samples of low-redshift galaxies confirm their prevalence \citep[e.g.,][]{Mullaney13, Woo16}, and studies at high redshift suggest they are equally common during the epochs of peak star formation and SMBH accretion \citep{Perna15, Brusa15, Zakamska16, Leung17} and into the epoch of reionization \citep{2022Natur.605..244B,2026Natur.653..368L}. Correlations between outflow properties and bolometric AGN luminosity \citep[e.g.,][]{Sturm11, Veilleux13, Rupke13, Sun17, Fiore17} suggest that AGN feedback may be most effective in quasars. Despite this, the properties and physics of AGN-driven outflows remain poorly understood \citep[see][for recent reviews]{Harrison18, Veilleux20}.
  
Low-redshift quasars are therefore excellent laboratories for studying large-scale, quasar-mode AGN-driven outflows and their effects on host galaxies. At $z\sim0.1$, they are the nearest SMBHs approaching the high accretion rates ($\dot{M} > 0.1 \dot{M}_{\mathrm{Edd}}$) seen in luminous AGN at high redshift, making them uniquely accessible for detailed study.

An open question in determining the physical properties of quasar-mode AGN-driven outflows is the extent of the outflowing gas. 
Deep optical integral field spectroscopy (IFS) of several of the nearest quasars show high-velocity, large-scale outflows in these systems, as well as correlations between kpc-scale outflows and SMBH or host galaxy properties \citep{Rupke11, Rupke13, Liu13, Harrison14, Husemann14, McElroy15, Rupke15, Karouzos16, Bae17, Rupke17, Kang18}. 

Some studies of quasar outflows in \oiii \ emission, using narrowband imaging coupled with longslit spectroscopy, found relatively small sizes with radii $\lesssim$1 kpc \citep[e.g.,][]{Fischer18, 2018MNRAS.478.2438S, 2021MNRAS.500.1491T}. Other longslit studies have found larger radial extents of $\sim$1-8 kpc \citep{Sun17}.  Past IFS studies typically detected outflows to larger radii \citep[$\sim$1-8 kpc,][]{Bae17, Kang18}, with some papers finding extents up to $\sim$20 kpc \citep{Liu13,Ulivi24}. IFS datasets from Gemini/GMOS, in particular, have often found outflows that fill the FOV, such that the radial extents measured were lower limits, particularly for low redshift luminous AGN at $z<0.1$ \citep{Harrison14,Rupke15,Rupke17,Kang18}. The sampling and FOV of Gemini---0\farcs1 and 3\farcs0 square, respectively---limit its sensitivity to low surface-brightness, extended emission, as may be found in a large-radius galactic wind.

One of the key open questions regarding the physical properties of quasar-mode AGN-driven outflows is the spatial extent of the outflowing gas. While results from many of these studies suggest wind sizes comparable to the disk scale lengths of their host galaxies, the precise physical extent remains uncertain, yet it is crucial for estimating mass and energy outflow rates and understanding feedback on the host galaxy.  Additionally, the extent of the outflowing gas is relevant for determining the role of AGN winds in shaping the circumgalactic medium (CGM), which 
is known to be thick with gas clouds \citep[e.g.,][]{Tumlinson13, Werk14} that may have been lifted by AGN-driven winds out of the host galaxy and into the surrounding halo.

Many of the current observational results on the radial extents of quasar outflows are, in
principle, lower limits, as they are limited by instrument sensitivity to low surface brightness emission and by the FOV of existing IFS. Instruments such as 
MUSE on the VLT \citep{Bacon10} and KCWI on Keck \citep{Morrissey18} are useful for this science,
 as they have an order of magnitude greater sensitivity as well as substantially larger fields of view.
In general, IFS is a highly valuable technique for
separating outflow structures from rotating gas disks and probing
their structure \citep[e.g.,][]{Rupke13,Bianchin26} and is the most reliable ways to simultaneously constrain the presence, scale and energetics of AGN-driven winds at all redshifts.

In this paper we present a KCWI survey of high-velocity outflows in seven low-redshift quasars (six of which are Type 2 AGN, and one of which is a narrow line Type 1 AGN)
to address a 
a key discrepancy raised by past studies: are wind sizes of order 1 kpc or 10 kpc?
We have selected Type 2 radio-quiet quasars and one narrow line Type 1 quasar with redshifts
$z = 0.04-0.12$ and $L_\mathrm{AGN} > 10^{45}$ erg~s$^{-1}$. 
While several of these luminous AGN have been previously observed with
IFS data \citep{Rupke11, Rupke13, Rupke15,  Harrison14, Rupke17,Kang18, Perna21,Ulivi24}, previous observations were often limited by small fields of view and/or relatively low sensitivities.

With these data we are able to measure the extent, velocity, and excitation of the outflowing gas, separating the emission from the outflow from host galaxy and AGN emission. 
Masses, momenta, and energies are estimated from line fluxes, velocities, and spatial extents. These masses, momenta, and energies ultimately determine how effective quasar winds are at creating feedback, impacting star formation, and influencing the CGM of their host galaxies.

The structure of the paper is as follows: in \S\ref{sec:obs-red} we describe the observations, data, and spectral fitting of the continuum and line emission. In \S\ref{sec:indiv-outflows} we present the results for the emission in each quasar individually, showing maps of the morphology and extent of the emission (typically \oiii) in each kinematic component, radial surface brightness profiles and spatial kinematic maps of the velocity centroid and width for each component, and example spectra and fits.  In \S\ref{sec:derived-properties} we estimate outflow physical properties such as ionized gas mass, mass outflow rate, kinetic energy and energy outflow rate, and momentum flux for each kinematic component in each quasar.  In \S\ref{sec:discuss}
we discuss the outflow extents measured here, as well as the incidence of detections of multiple outflows within individual quasar hosts, and we compare the properties of the distinct outflows detected within the same sources as a function of the outflow velocity. 
Lastly, in \S\ref{sec:concl} we present a summary and conclusions of our results.
Throughout this study we adopt the following cosmology: $\Omega_{\rm{m}}=0.315$, $\Omega_{\Lambda}=0.685$,  $h=0.674$.


\input{targets_table}  

\section{Observations, Data, and Spectral Fitting} \label{sec:obs-red}

\subsection{Sample Selection and Observations}

We observed seven low-redshift quasars ($L_\mathrm{AGN} > 10^{45}$ erg~s$^{-1}$) with the Keck Cosmic Web Imager \citep[KCWI,][]{Morrissey18} on the Keck II telescope on November 6, 2018. The quasars have $z = 0.04-0.12$; possess known high-velocity outflows; are radio-quiet (with one exception); and are Type 2 (with one exception, a narrow line Type 1). We prioritized quasars with measurements of outflow sizes from either IFS \citep{Harrison14, Rupke17, Kang18, Perna21, Ulivi24} or long-slit spectroscopy \citep{2013MNRAS.432..138R, Sun17, 2018MNRAS.478.2438S, Fischer18}. Due to the selection by previous outflow measurements from various sources, the quasars have heterogeneous origin, with a combination of infrared and optical selection (see Section \ref{sec:indiv-outflows} for further details on individual objects).

We observed the sample with the KCWI medium slicer; the blue low-dispersion (BL) grating, tuned to a central wavelength of 4500 \AA; the KBlue filter; and 2$\times$2 spatial binning. This configuration results in a spaxel size of 0.29$\arcsec$ $\times$0.69$\arcsec$ and a field of view of 16$\arcsec$ $\times$ 20$\arcsec$ per pointing.
The BL grating has a spectral resolution of $R=$1800, and the observed wavelength range was $\sim$3600-5500 \AA. We took exposures of 10 minutes each and dithered half a slicer (0.35\arcsec) between exposures. We took additional exposures of 1 minute each for each target, to account for any saturation in the central spaxel(s) in the 10 minute exposures. Total exposure times are listed in Table 1 and varied from $\sim$40 to 90 minutes. All targets were observed at a position angle of 0 degrees. 
The sky was photometric, and the seeing was $\sim$0.6\arcsec.

\subsection{Data Reduction}

We reduced the data using the IDL version of the KCWI Data Reduction Pipeline (DRP v1.2.1) and the IFSRED IDL library \citep{Rupke14a}. Sky subtraction was performed using the DRP with manual selection of a sky mask region within each pointing. A correction was made for the saturated spaxels near the nucleus, using the 1 minute exposures. The standard star BD+28411 was observed to provide flux calibration. Wavelengths are given in air.

Following these pipeline stages, we resampled the data onto 0.29$\arcsec$ $\times$ 0.29$\arcsec$ spaxel grids using the routine IFSR\_KCWIRESAMPLE. The resulting data cubes typically have dimensions of $\sim$80 spaxels on a side, corresponding to 23$\arcsec$, though the exact FOV depends on the number of pointings observed per target.

\subsection{Spectral Fitting}

In this paper we aim to determine the physical extents and spatially-resolved kinematics of outflows in the observed quasars.  While some papers in the relevant literature utilize non-parametric fitting of AGN spectra to identify and characterize outflows, here we retain the multiple-Gaussian fits to the emission lines. This allows for the possibility of identifying multiple, kinematically-distinct outflows in the same quasar. As described below, we use spatial maps of the kinematic properties in each kinematic component and perform extensive checks of the spectral fits to verify our results.  We utilize the publicly-available IFSFIT code \citep{Rupke14b} as described below to fit the data.

%
\begin{figure}[t]
    \centering
    \includegraphics[width=0.45\textwidth]{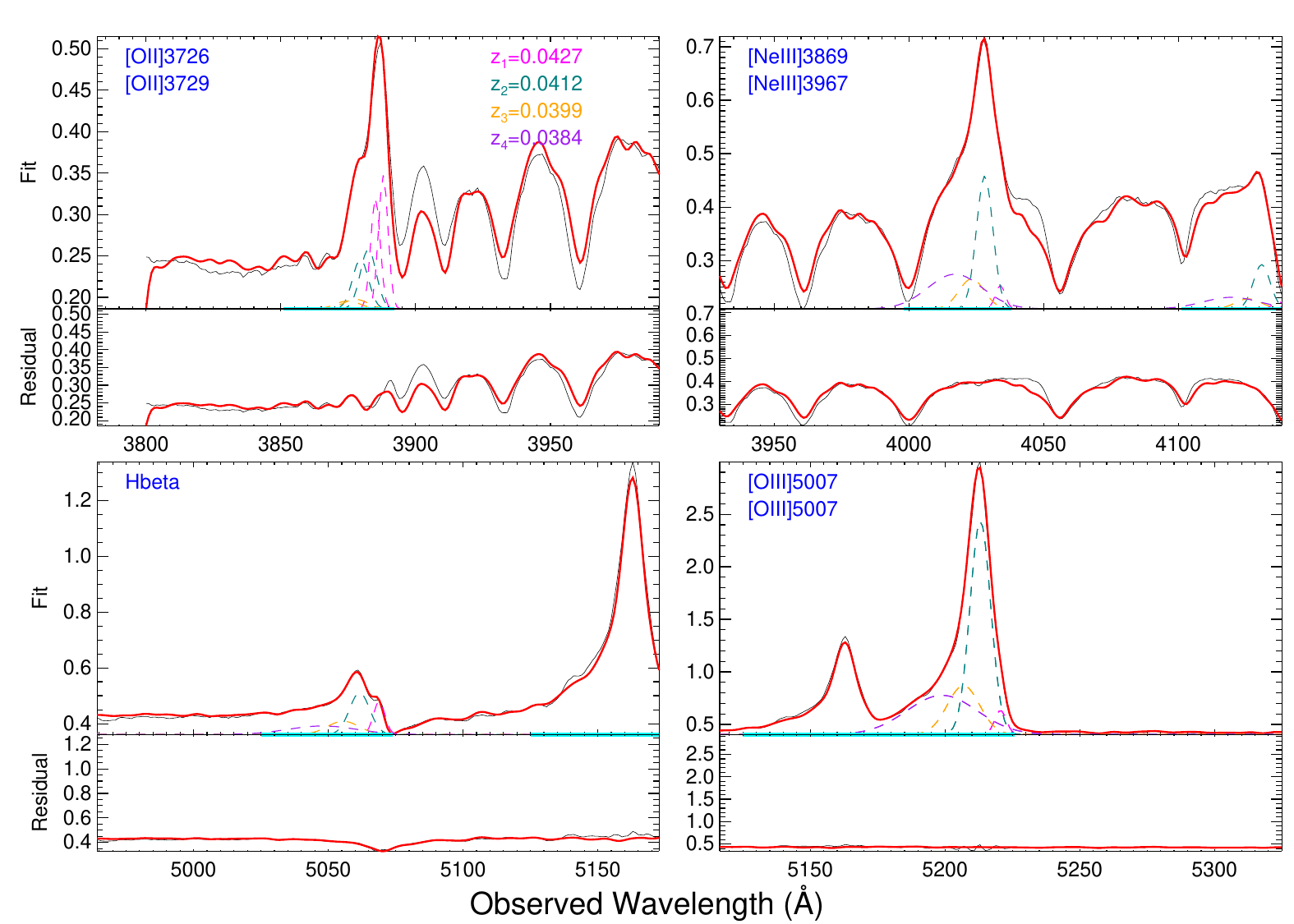}
    \caption{Example spectral fitting: Combined stellar and AGN continuum and emission line fits (in red, upper panels) showing \oii, \neiii, \hb, and \oiii \ in the central, brightest spaxel in the F05189-2524 mosaic. The data are shown in black. The lower panels show the continuum fits after subtracting the line emission, shown with dotted pink lines, and cyan regions indicate the wavelengths used to identify and fit the emission lines.}
    \label{fig:IFSFIT_example}
\end{figure}

\begin{figure*}[t]
\centering
\includegraphics[width=0.9\textwidth]{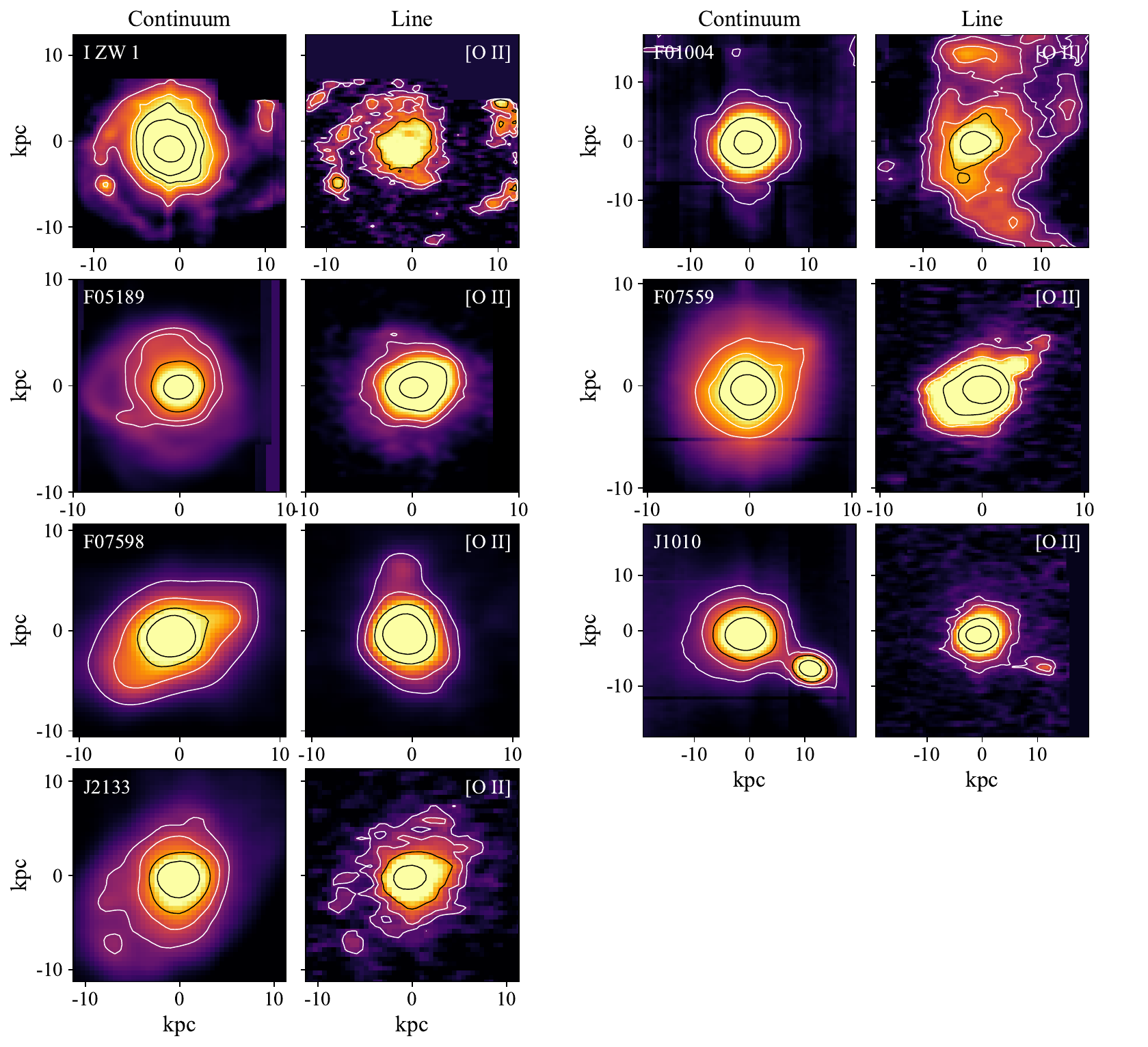}
\caption{KCWI maps of the continuum (left panels, integrated over the observed wavelength range 3700-5300 \AA) and \oii \ line emission (right panels) for each quasar in the observed sample.  In all panels north is up.  The continuum maps are typically dominated by light from the AGN, with some contribution from stars in the host galaxy, while the line emission maps often trace structure further out in the host galaxy from emission from HII regions.  Several of the quasar host galaxies are known to be late stage major mergers (F05189, F01004), while others show clear spiral structure (I Zw 1).}
\label{fig:cont_line_maps}
\end{figure*}

\begin{figure*}[t]
\centering
\includegraphics[width=0.8\textwidth]{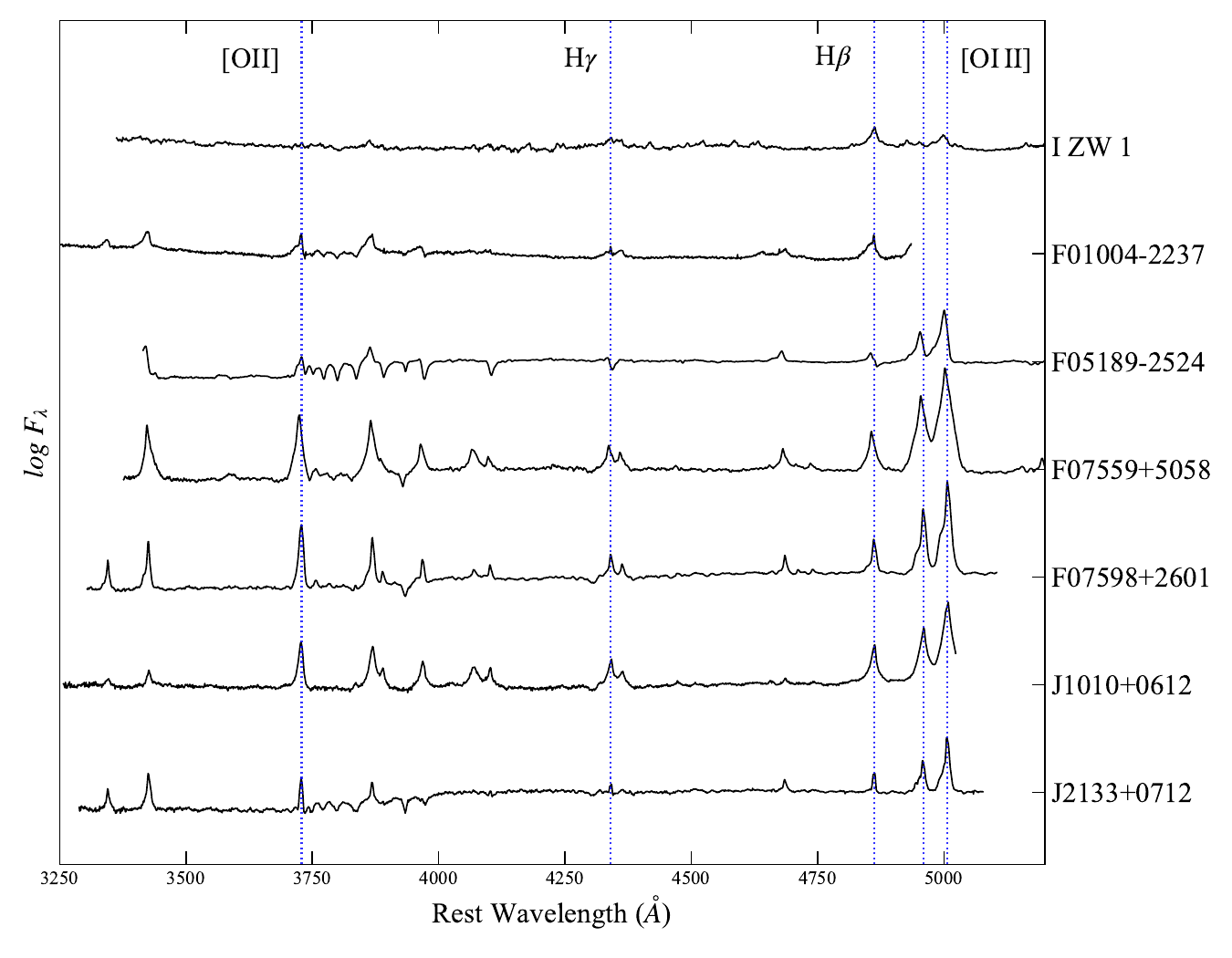}
\caption{Spectra of the log of the flux in the brightest spaxel in the KCWI mosaic for each quasar in this sample, shown as a function of restframe wavelength.  The spectra are offset for clarity.  The observed wavelength coverage includes \oiii \ and \hb \ emission  for all of the quasars except for F01004, which has \hb \ but not \oiii \ 5007 \AA \ coverage.
}
\label{fig:spectra}
\end{figure*}

\subsubsection{IFSFIT}

For all of the quasars in this sample we use the IFSFIT library \citep{Rupke14b} in IDL to model the spectrum in each spaxel. This code allows for simultaneous fits to the stellar continuum emission and line emission, with an option to separate nuclear unresolved AGN light (see \citealt{Rupke17} for details). This package incorporates PPXF \citep{Cappellari12} to fit the stellar continuum and MPFIT \citep{Markwardt09} to fit Gaussian profiles to the emission lines. IFSFIT first masks emission
BPASS \citep{Stanway18, Byrne22} high-resolution stellar population synthesis model assuming solar metallicity, with additional Legendre polynomials allowed to account for residuals from imperfect data calibration such as scattered light or spatially-varying residual sky background.

In fitting the emission lines, we typically fit the \oii \ 3726\AA, \oii \ 3729\AA, H$\beta$, \oiii \ 4959\AA, \oiii \ 5007\AA, [\textrm{N}~\textsc{i}] 5198\AA, and [\textrm{N}~\textsc{i}] 5200\AA \ lines, depending on the redshift of the source and which lines are observed.  The minimum allowed value of the velocity dispersion (\vsig) in the emission lines fits is 50 \kms. We vary the maximum allowed number of Gaussians from two to four for each quasar, where each Gaussian corresponds to a different kinematic component; only those kinematic components fit with high S/N ($>2.5$) are retained.  The kinematics of all of the  emission lines are tied so that multiple lines are fit simultaneously and share the same kinematic parameters by design, though not all Gaussian components are required for every emission line.

For each spaxel and each kinematic component we measure the emission line velocity dispersion (\vsig) and the  velocity relative to the systemic redshift (listed in Table 1) at which 2\% ($v_{02}$), 50\% (\vcen), and 98\% ($v_{98}$) of the line flux accumulates integrating from red (positive velocities) to blue (negative velocities) across the line profile. Here \vcen \ is the median velocity in each profile, while $v_{02}$ and $v_{98}$ measure the maximum redshifted and blueshifted velocity wings, respectively \citep{Rupke13}.
We define the maximum velocity, $v_{max}$, to be $v_{02}$ for redshifted emission and $v_{98}$ for blueshifted emission.

Model line profiles are
convolved with the spectral resolution before fitting, such that the instrumental resolution is taken into account.  As the \oii \
doublet is unresolved in our observations,  the \oii \ 3729/3726 \AA \ flux
ratio is fixed to 1.2, corresponding to an electron density of 400
$cm^{-3}$ \citep{Pradhan06}.  Emission lines with a significance 
less than 2.5$\sigma$ in their flux are set to zero in each spaxel. From these fits we then 
create flux and kinematics spatial maps for multiple emission lines for each Gaussian component for each AGN. IFSFIT fits the spectra independently in each spaxel, without using information about fits in nearby spaxels.
 
For spaxels with multiple kinematic components fit, following \citet{Rupke17} the components were sorted
into distinct maps by wavelength (\vcen) or velocity dispersion (\vsig). For a given galaxy, the first component (c1) has either a smaller \vcen \ or a smaller velocity dispersion than the second component (c2). Several quasars have three components, while one (F05189) has four components.  The choice of sorting by central velocity or velocity dispersion was applied to the entire cube.
The choice was based on whichever method produced the most coherent kinematic signal in spatial maps of \vcen \ and \vsig \ across components.
Details of how the components are defined for each quasar are given in the section below.

We ran IFSFIT for each AGN using a systemic redshift published in the literature and updated the output to define the redshift such that the median \vcen \ across all spaxels is defined to be 0 for the first kinematic component, associated with the host galaxy (details are given below for each quasar). The updated systemic redshift is consistent with the stellar redshift fit by IFSFIT for each AGN.  Outflow components are typically blueshifted and have higher \vsig \ values than the first component, though some outflow components are redshifted, and the \vcen values of the emission both from the host galaxy (c1) and outflow(s) can spatially vary across the galaxy.

Figure \ref{fig:IFSFIT_example} shows an example  of fits using IFSFIT for the spectrum in the central, brightest spaxel in the mosaic for F05189-2524.  This figure shows combined continuum and line emission fits, where the red line shows the total fit.  In this quasar four distinct kinematic components are detected in the emission lines; the first component, corresponding to emission with \vcen \ near 0 \kms, is best traced by the \oii \ emission, while the outflowing gas observed at blueshifted velocities is best traced by  \oiii \ emission and is also detected in the \neiii \ and \hb \ emission profiles. As all of the emission lines are fit simultaneously, we are typically able to obtain robust constraints on the first kinematic component from the \oii \ emission, while the higher ionization lines dominate the outflow components, as seen in this example. The \hb \ emission line often has substantial contributions from both the first component and other kinematic components, arising from gas both in the host galaxy and in the outflow(s). 

For the luminous AGN in our sample the continuum emission is due to light both from the AGN and the stars in the host galaxy.  We do not attempt here to separate the stellar and AGN contributions to the continuum fits, and in this paper we do not utilize the option available in IFSFIT to fit the spatially-unresolved quasar light separately. We tested this option (which is used extensively in \citet{Rupke17} on integral field data from Gemini-N/GMOS of outflows in Type 1 broad-line quasars) and found that with Keck/KCWI data residual low level scattered light in the reduced mosaic prevented a robust subtraction of the unresolved quasar light. In detail, there are small differences in the shape of the AGN spectrum that vary spatially across the mosaic, which prevents clean subtraction of the unresolved AGN light separately.  However, performing continuum fits using a combination of stellar models and Legendre polynomials results in excellent fits.

\section{Outflow Maps, Radial Extents, and Kinematics}
\label{sec:indiv-outflows}

Spatially-resolved maps of each quasar in the sample are shown in Figure \ref{fig:cont_line_maps} for both the integrated flux from the continuum emission (left panels) and line emission (right panels) from \oii.  The continuum flux maps are dominated by the bright quasar light in the center of the galaxy, which in two sources is sufficiently bright to cause diffraction spikes or scattered light: I Zw 1 and F07559. The continuum maps also include contributions from the stellar continuum emission in the host galaxy.  The line emission maps typically trace emission from HII regions further out in the host galaxy and as such these maps are not as centrally concentrated as the continuum maps.  In several cases clear structure can be seen in the host galaxy such as tidal features or spiral structure; many of the host galaxies in this sample are ULIRGs.
Figure \ref{fig:spectra} displays the spectra of the brightest spaxel in each mosaic, dominated by the quasar light.

In this section we present the spectral fitting results for each quasar, 
detailing the spatially-resolved properties of the host galaxy and outflow emission and utilizing our ability to identify and separate multiple kinematic components. 
Table 2 lists the number of kinematic components identified and the velocities and radial extent of the gas in each component. 
Table 3 lists the observed and dust-corrected \oiii \ 5007 \AA \ luminosities for the total emission summed over all spaxel, as well as the emission in each kinematic component. 
For each quasar we provide composite figures illustrating the kinematics, surface brightness, and spatial extent of the emission for a given  line (typically \oiii), as well as spatial maps of the \oii / \oiii \ line ratio and example spectra and fits in different locations around the source.  The details of the morphologies, extents, line ratios, and kinematics are discussed below for each quasar.

\subsection{I Zw 1}

\begin{figure*}[t!]
\centering
\includegraphics[width=0.97\textwidth, trim=0.5cm 3cm 1.5cm 0.5cm, clip]{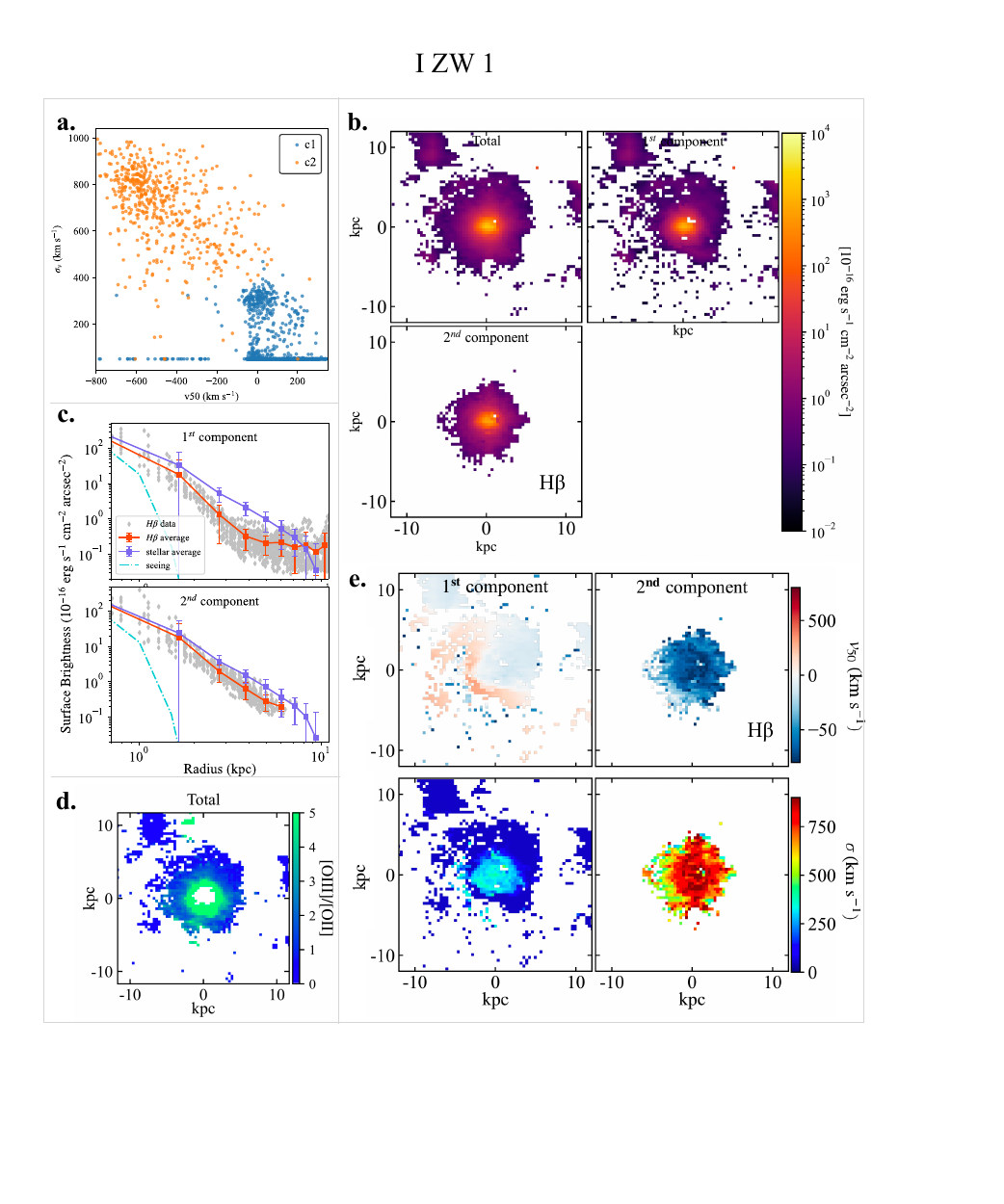}
\caption{Results for I Zw 1: Panel a (upper left) shows the central velocity, $v_{50}$, and velocity dispersion, $\sigma$, of each kinematic component fit for each spaxel in the I Zw 1 mosaic. Two kinematic components are detected in this quasar; `c1' is the first component, shown in blue, while `c2' is the second component, shown in orange. Panel b (upper right) shows spatial surface brightness maps for the total \hb \ emission and the emission for each kinematic component. Panel c (middle left) shows azimuthally-averaged radial surface brightness profiles for the \hb \ emission for each kinematic component (red lines show the average while grey points show the values for each spaxel), while the blue lines show the average stellar radial profile in the host galaxy, normalized to the peak line emission surface brightness in that profile.  The seeing profile is also shown normalized to the line emission in each subpanel. Errors are the standard deviation from the mean.  Panel d (lower left) shows a spatial map of the \oiii \ 5007 \AA / \oii \ doublet line ratio for total emission (summing over both kinematic components). Panel e (lower right) shows spatial maps of the \hb \ kinematics in each kinematic component, where the top row shows the central velocity, \vcen, and the bottom row shows the velocity dispersion, \vsig.}
\label{fig:IZw1-1}
\end{figure*}

\begin{figure*}[t!]
\ContinuedFloat
\centering
\includegraphics[width=0.75\textwidth, trim=0.5cm 12.4cm 1.5cm 0.5cm, clip]{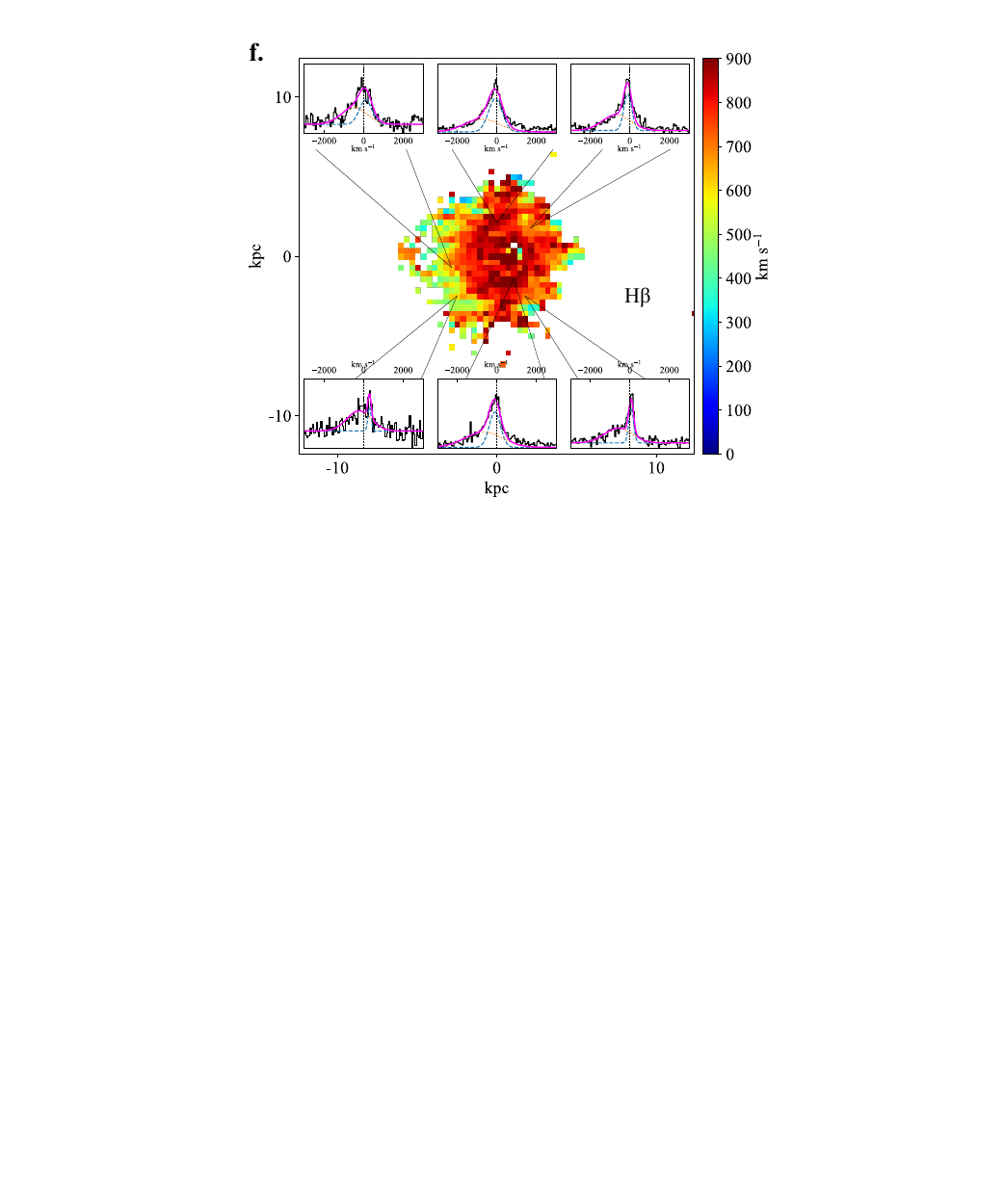}
\caption{Results for I Zw 1 (continued): Panel f shows the velocity dispersion, \vsig, of the second component in the large central map with velocity profiles for the \hb \ emission shown for 6 spaxels as marked. For the velocity profiles the black line shows the continuum-subtracted data, the pink lines shows the total fit, and the two separate Gaussian fits are shown with blue dashed and orange dotted lines. The blue fit corresponds to the first component, c1, and the orange fit corresponds to the second component, c2.}
\label{fig:IZw1-2}
\end{figure*}

\input{outflow_properties_table.tex} 

\input{oiii_luminosities.tex} 

I Zw 1 is a known narrow line Seyfert 1, one of the original Palomar-Green quasars \citep{1986ApJS...61..305G}, with \hb \ emission that is broader than typically observed in NLAGN. \citet{Rupke17} observe this quasar with Gemini GMOS IFS data but do not detect an ionized outflow; only a neutral outflow observed in Na D absorption is detected. \citet{Perna21} publish MUSE IFS data on I Zw 1 and report a blueshifted outflow with a maximum velocity observed in \oiii \ of $W_{80}=2190$ \kms. Using the same data, \citep{2022ApJ...935...72M} report a radius of 17.6\,kpc for the NLR.

The host galaxy of I Zw 1 is a face-on spiral, and the continuum map for this source (Figure \ref{fig:cont_line_maps}) shows emission from the AGN as well as the spiral arms of the host galaxy, which are seen clearly in the line emission map showing \oii \ emission. The \hb \ emission is stronger in this source than the \oiii \ emission (Figure \ref{fig:spectra}), unlike for the rest of the quasars in this sample in which the \oiii \ emission is the strongest. 

We use IFSFIT to fit the \oii, \hg, \hb, and \oiii \ emission in this AGN and find two kinematic components. A plot of the \vcen \ versus \vsig \ values fit in each spaxel (panel a of Figure \ref{fig:IZw1-1}) shows two distinct groupings of fits: a cluster near the systemic redshift with relatively low \vsig ($<400$ \kms) and a large number of blueshifted fits up to \vcen$=-800$ \kms \ spanning \vsig$\sim400-1000$ \kms.  When assigning Gaussian profiles to kinematic components, when two components are fit in a given spaxel the component with higher \vsig \ is assigned to the second kinematic component (c2). If only one Gaussian profile is fit in a given spaxel, it is assigned to the first component (c1) if \vsig$<380$ \kms \ and to the second kinematic component (c2) if \vsig$>380$ \kms. This results in the fits being assigned with the colors as shown in panel a of Figure \ref{fig:IZw1-1}.

For I Zw 1 we show results for the \hb \ emission rather than \oiii, as \hb \ is the strongest line in this source and both kinematic components are well fit across the full extent of the mosaic. The \oii \ emission is dominated by the first kinematic component, while \oiii \ is dominated by the second; since the kinematics of all lines are fit simultaneously, the spatial maps for the second component in \oiii \ look very similar to those shown here for \hb.

Panel b in Figure \ref{fig:IZw1-1} shows surface brightness maps of the \hb emission in each kinematic component, as well as for the total emission summed over both components.  Panel c shows the surface brightness radial profiles of the \hb \ emission both the first and second kinematic components, along with the (scaled) profiles of the stellar continuum fit and the seeing.

For this source the \oiii / \oii \ map (panel d of Figure \ref{fig:IZw1-1}) is shown only for total flux and not shown for the two kinematic components separately, as the second kinematic component is not typically detected in the \oii \ emission, which has a relatively low $S/N$.  Using the integrated flux in each line for this line ratio, the values are high near the center of the galaxy at the location of the AGN and fall to lower values at larger radius in the host galaxy.

The spatial maps of the first kinematic component (panels b and e) show that it is dominated by light from the host galaxy, with a median \vsig$=50$ \kms \ and a radial extent of $>10$ kpc, filling the field of view of the mosaic.  The spatial kinematic maps (left column of panel e) show coherent \vcen \ with redshifted spiral structure to the southeast ($>150$ \kms) and blueshifted emission ($\sim-100$ \kms) to the northwest.  The \vsig \ values in this component are elevated in the center of the galaxy, near the AGN; this is due to this AGN being a  narrow line Seyfert 1, with broader wings in the \hb \ profile.

The second kinematic component for this quasar is blueshifted, with a median \vcen$=-543$ \kms \ and a high velocity dispersion with median \vsig$=732$ \kms. The radial extent of the blueshifted outflow is $\sim$5 kpc.  This outflow is observed along the line of sight with only blueshifted emission detected. The \vcen \ values are $\sim-600$ \kms \ across the bulk of the spatial map of the emission, decreasing to a lower blueshift of $\sim-300$ \kms \ to the southeast.
The \vsig \ values are $\sim$800 \kms \ for most of the spaxels, decreasing slightly to $\sim$600 \kms \ to the southeast.  

Example spectra and fits are shown in panel e of 
Figure \ref{fig:IZw1-1} for the \hb \ emission line. Here the blue dashed line corresponds to fits associated with the first component, while orange dotted lines show fits associated with the second component.  The median velocities and radial extent of each component are listed in Table 2, and Table 3 lists the \oiii \ luminosity in each kinematic component.  Here the dust-correction is derived from the Balmer decrement, as discussed in the next section below.

\subsection{F01004-2237}

\begin{figure*}[t!]
\centering
\includegraphics[width=0.97\textwidth, trim=0.5cm 3cm 1.5cm 0.5cm, clip]{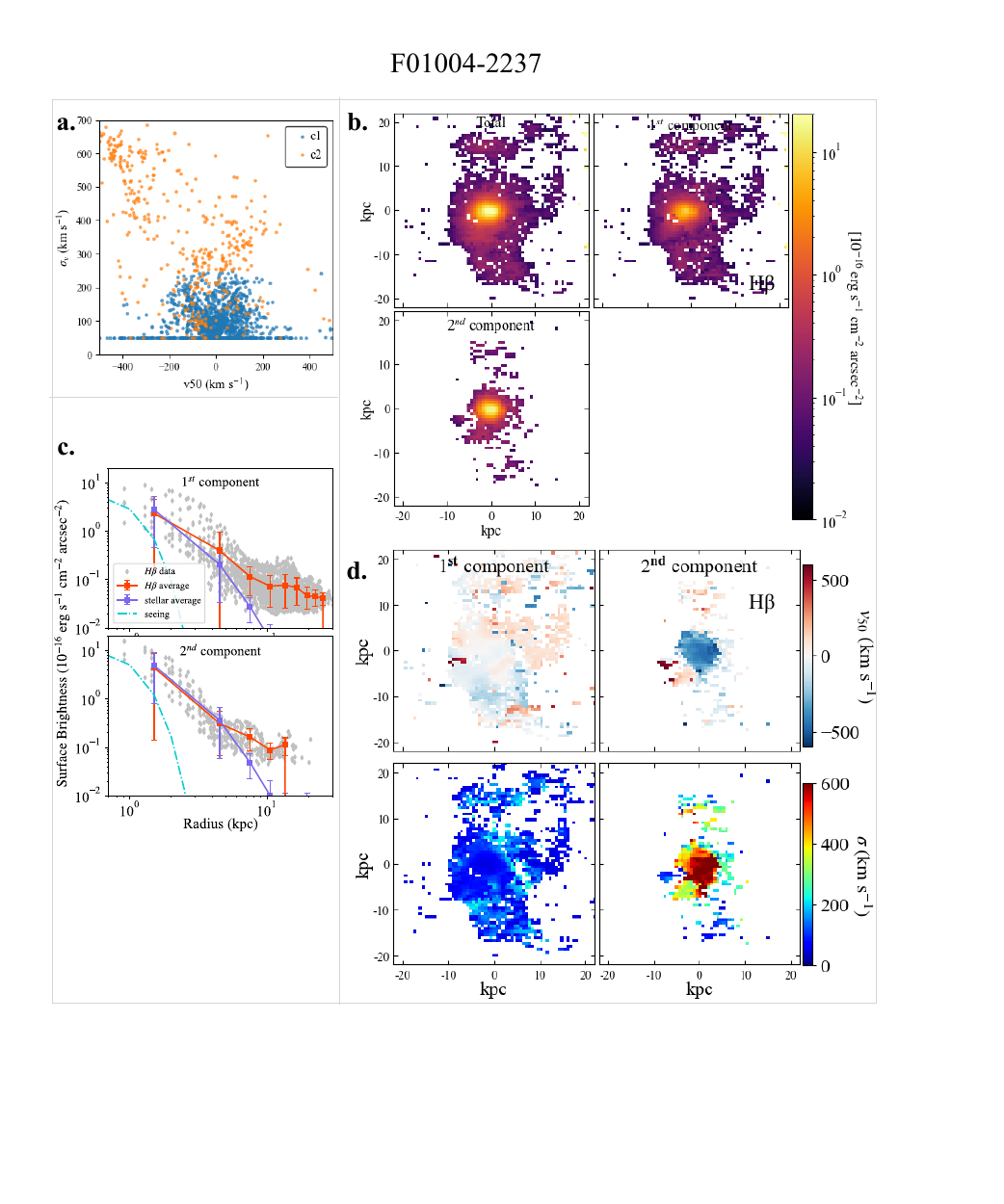}
\caption{Results for F01004-2237: panels show the same information as in Figure 4.  We do not observe the \oiii \ emission line for this quasar and therefore show the \hb \ emission maps.  No 
 \oiii \ 5007 \AA / \oii \ doublet line ratio  map is available for this source.}
\label{fig:F01004-1}
\end{figure*}

\begin{figure*}[t!]
\ContinuedFloat
\centering
\includegraphics[width=0.75\textwidth, trim=0.5cm 13cm 1.5cm 0.5cm, clip]{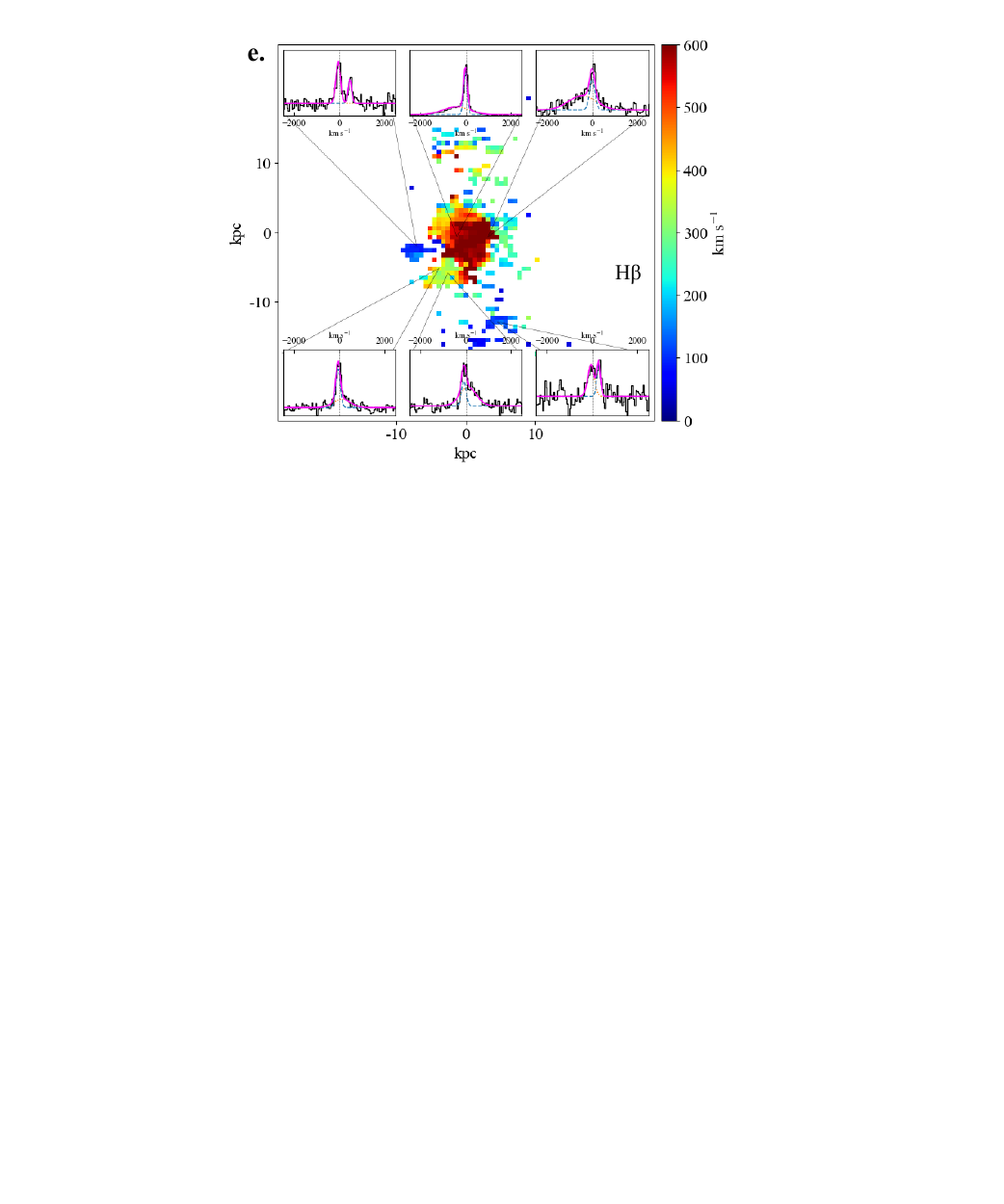}
\caption{Results for F01004-2237 (continued): panel e shows the same information as in Figure 4, here for the \hb \ emission in this quasar.}
\label{fig:F01004-2}
\end{figure*}

F01004$-$2237 (hereafter F01004) is an ultraluminous infrared galaxy (ULIRG; \citealt{1999ApJ...522..113V}). While the narrow emission-line components are ionized by starlight, a broad, highly blueshifted line wing in most optical lines is likely AGN-photoionized, pointing to a Type 2 quasar \citep{1999ApJ...522..113V,2010ApJ...709..884Y,2012ApJS..203....3S,2013MNRAS.432..138R}. Furthermore, the mid-IR emission of F01004 suggests a dominant AGN contribution \citep{2009ApJS..182..628V}. It has a UV broad absorption line (BAL) in \ion{O}{6} indicative of a high-velocity, unresolved quasar outflow \citep{2022ApJ...934..160L}. Longslit spectra show two outflow components in \oiii, with central velocities of -300 \kms \ and -1000 \kms\ \citep{2013MNRAS.432..138R,2018MNRAS.478.2438S}. The inferred outflow radius from HST spectroscopy is tiny, at only 111~pc \citep{2018MNRAS.478.2438S}, while ground-based data suggest 3~kpc \citep{2012ApJS..203....3S,2013MNRAS.432..138R}.

Notably, F01004 has spectral and photometric variability, in particular in its line emission \citep{2017NatAs...1E..61T,2021MNRAS.504.4377T,2024A&A...692A.262S}. This is attributed either to a partial tidal disruption event (TDE) showing multiple episodes \citep{2024A&A...692A.262S}, multiple single TDEs \citep{2017NatAs...1E..61T,2024A&A...692A.262S}, or AGN-related flaring \citep{2019NatAs...3..242T}. Relevant for this work, the \oiii\ broad emission lines show some variation due to light echoes, increasing by a factor of 1.4--1.8 from 2015 to 2018 \citep{2021MNRAS.504.4377T}.

The KCWI continuum emission map (Figure \ref{fig:cont_line_maps}) for this galaxy is dominated by light from the AGN, while the line emission map more clearly shows the host galaxy, a major merger with substantial emission on large scales showing tidal features such as loops and tails.  The \oii \ emission fills most of the field of view of the mosaic for this source.

F01004 has a slightly higher redshift than the rest of the quasars in this sample, such that the observed spectrum (Figure \ref{fig:spectra}) does not cover the \oiii \ emission used to trace outflows in the rest of the sample.  Here we use IFSFIT to fit the \oii, \hg, and \hb \ emission and find two kinematic components.  Results are shown in Figure \ref{fig:F01004-1} for the \hb \ emission, and this AGN is not included in Table 3, which lists the \oiii \ 5007 \AA \ luminosities, as that line is not covered for this source.

The \vcen-\vsig \ diagram (panel a of Figure \ref{fig:F01004-1}) shows a cluster of fits near the systemic redshift with low \vsig ($\lesssim$ 200 \kms, along with a fewer number of points spanning \vsig$\sim$300-700 \kms, most of which are blueshifted up to $-500$ \kms.
When assigning Gaussian profiles to kinematic components, when two components are fit in a given spaxel the component with higher \vsig \ is assigned to the second kinematic component (c2). If only one Gaussian profile is fit in a given spaxel, it is assigned to the first component (c1) if \vsig$<250$ \kms \ and to the second kinematic component (c2) if \vsig$>250$ \kms. This results in the fits being assigned with the colors as shown in panel a of Figure \ref{fig:F01004-1}.

The gas in the first kinematic component has a median \vsig=87 \kms \ and extends to a radius of $>$18 kpc. The spatial kinematic maps for this component (left column of panel d) show a highly complex kinematic structure in \vcen, reflecting the ongoing merger in this galaxy. No spatial maps of the \oiii/\oii \ ratio are available for this source, given the lack of coverage of the \oiii \ emission.

The gas in the second kinematic component is predominantly blueshifted, with \vcen \ $sim$-400 \kms \ in the center of the source, though \vcen \ is redshifted in the southeast by $\sim$130 \kms.  
The median \vsig \ of this outflow component across all spaxels is 330 \kms, while in the center of the galaxy this increases to \vsig$\sim$600 \kms. In the southeast where the emission is redshifted the velocity dispersion is lower with  \vsig$\sim$300 \kms.

The second kinematic component has an observed radial extent of $\sim$8 kpc.  The spaxels at larger radius (r$>$10 kpc) may be associated with the host galaxy merger, as \vsig \ is lower on the whole for these spaxels, which are spatially disjoint from the high velocity dispersion gas observed in the center of the galaxy.  Here we appear to be observing the front and back sides of an outflow in the blueshifted and redshifted emission at high velocity dispersion.

\subsection{F05189-2524}

\begin{figure*}[t]
\centering
\includegraphics[width=0.97\textwidth, trim=0.5cm 3cm 1.5cm 0.5cm, clip]{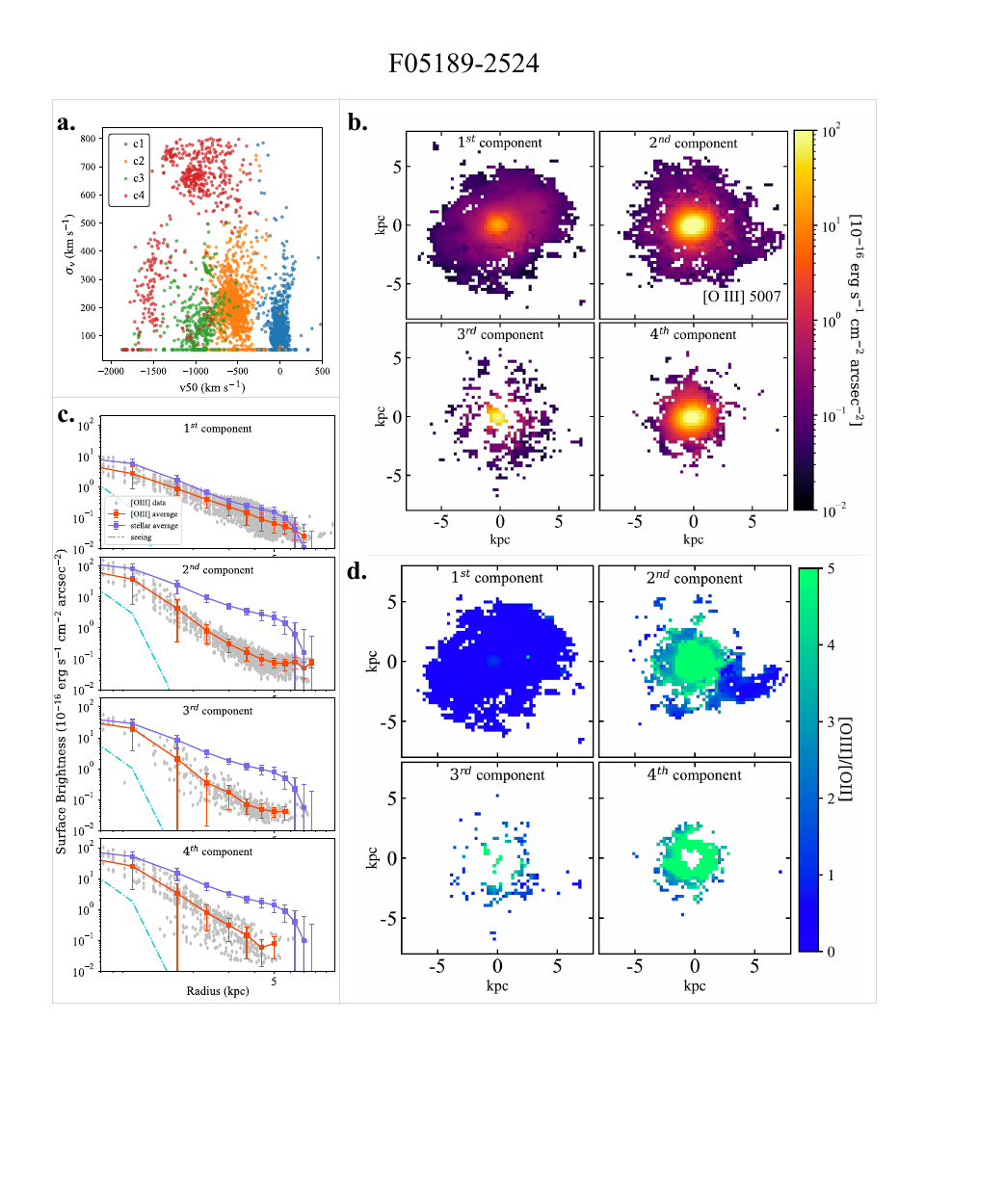}
\caption{Results for F05189-2524: Panel a (upper left) shows the central velocity, $v_{50}$, and velocity dispersion, $\sigma$, of each kinematic component fit for each spaxel in the F05189 mosaic. Four kinematic components are detected in this quasar, shown with different colored points ('c1' is the first component, etc.). Panel b (upper right) shows spatial surface brightness maps for the \oiii \ 5007 \AA \ emission for each kinematic component. Panel c (lower left) shows azimuthally-averaged radial surface brightness profiles for  the \oiii \ 5007 \AA \ emission for each kinematic component; here the seeing profile has been smoothed by the same amount as the data to account for the boxcar smoothing of the mosaic. Panel d (lower right) shows spatial maps of the \oiii \ 5007 \AA / \oii \ doublet line ratios for each kinematic component.}
\label{fig:F05189-1}
\end{figure*}

\begin{figure*}[t]
\ContinuedFloat
\centering
\includegraphics[width=0.75\textwidth, trim=0.5cm 3.cm 0.1cm 0.5cm, clip]{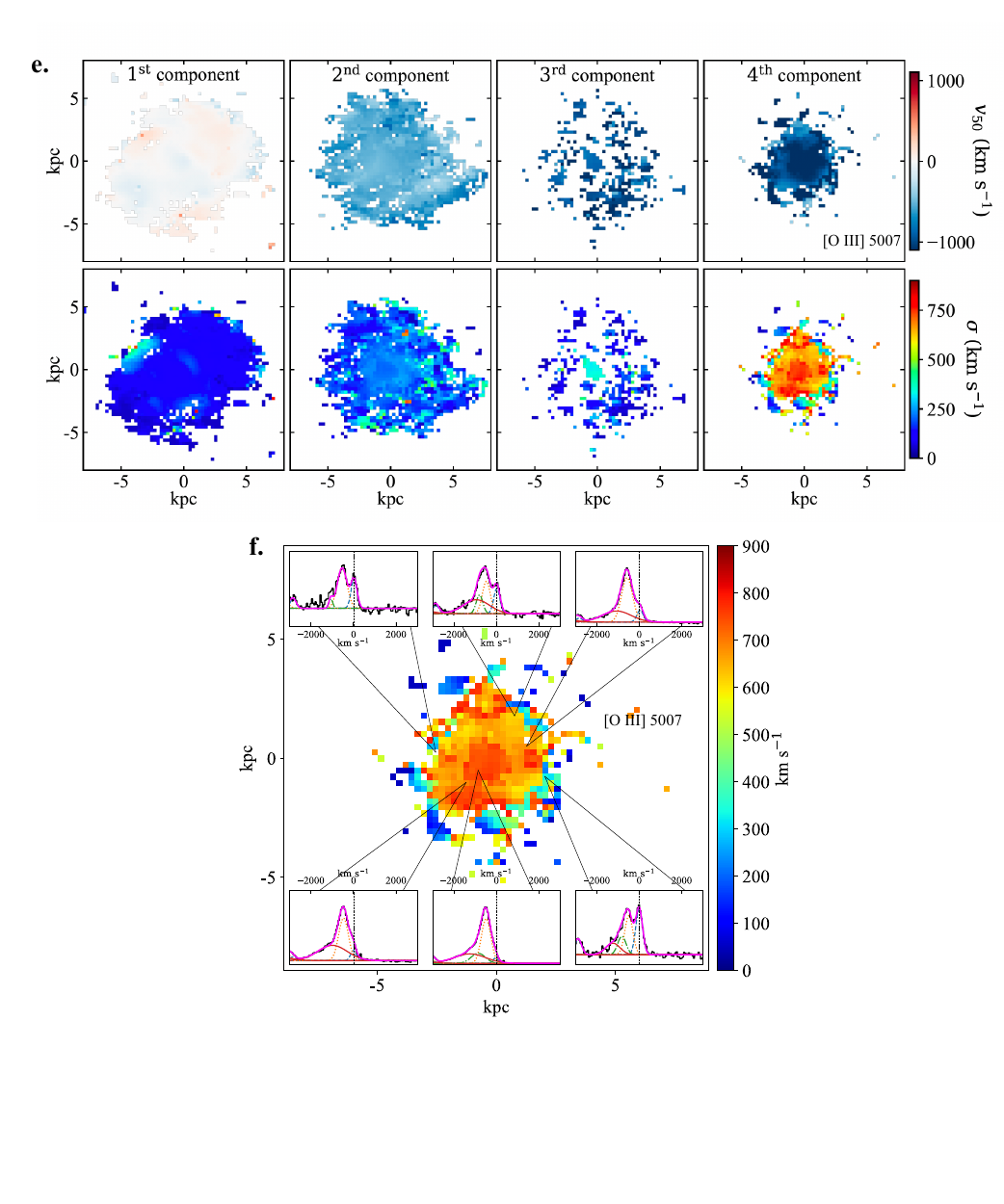}
\caption{Results for F05189-2524 (continued): Panel e (upper row) shows spatial maps of the \oiii \ 5007 \AA \ kinematics in each component, where the top row shows the central velocity, \vcen, and the bottom row shows the velocity dispersion, \vsig. Panel f (lower panel) shows the velocity dispersion, \vsig, of the fourth component in the central map with velocity profiles for the \oiii \ 5007 \AA \ emission shown for 6 spaxels as marked. }
\label{fig:F05189-2}
\end{figure*}

F05189-2524 (hereafter F05189) is a ULIRG at $z = 0.0424$ which hosts an AGN with a bolometric luminosity of $\log(L_{\mathrm{bol}}/L_{\odot}) = 12.2$, classifying it as a quasar \citep{Rupke17}. This galaxy is a known late-stage major merger \citep{2002ApJS..143..315V}, and there are clear tidal features observed in the KCWI continuum emission maps as multiple loops and shells of stellar material (Figure \ref{fig:cont_line_maps}). 

F05189 is one of the nearest and brightest ULIRGs, making it a prime target for detailed studies of the molecular \citep[e.g.][]{Gonzalez17}, neutral \citep[e.g.][]{Rupke05,Teng13}, warm ionized (e.g. \citealp{Bellocchi13}), and ultra-fast \citep{2019ApJ...887...69S,2025ApJ...993L..53N} outflows observed in this galaxy.  The optical spectrum is that of a narrow line AGN, though it has broad emission line components with FWHM$\sim$1700~\kms\ observed in recombination lines in the near infrared \citep{Veilleux99, 2001A&A...368...44S, 2026ApJS..282...68G}. This broad, blueshifted component could represent a classical broad line region (BLR) around an active galactic nucleus. It is also consistent with the strongly blueshifted outflow seen in strong optical lines, including in the forbidden lines not emitted by the BLR.

\citet{Rupke17} calculate that the mass outflow rate of the ionized, neutral, and molecular phases are 0.7, 26, and 73\%\ of the total, respectively. All of these gas phases have an observed high velocity and large-scale outflow. Up to two kinematic outflow components are observed in the neutral and molecular phases. Only one outflow component is observed in the resolved ionized phase in \citet{Rupke17}. However, the unresolved nuclear spectrum, which contains multiple outflow components, was first subtracted from the data cube. The radial extent of the ionized outflow is measured as 3 kpc in \citet{Rupke17}, though that data is limited by the GMOS field of view. They measure the mean and maximum value (over spaxels) of \vcen\ of the ionized gas to be 420 and 990 \kms, respectively.
MUSE observations of F05189 in \citet{Perna21} report a maximum outflow velocity observed in \oiii \ of $W_{80}=1860$ \kms, but no radial extent is reported.

For this source we use IFSFIT to fit the \oii, \neiii, \hg, \hb, \heii, and \oiii \ emission lines. We identify four distinct kinematic components in F05189, three of which are blueshifted (see panels a and e of Figure \ref{fig:F05189-1}). Of all of the quasars in our sample, this is the only AGN that has a total of four kinematic components clearly detected over a number of spaxels.  To increase the $S/N$ in the outer regions of this source we ran IFSFIT on a mosaic that was spatially smoothed by a 2$\times$2 boxcar, as on the full resolution mosaic we detected multiple kinematic components but the $S/N$ was too low at larger radius to robustly detect the fourth component. Fitting on the smoothed mosaic improved our ability to detect all four components across the FOV.  We also ran IFSFIT allowing a maximum of three kinematic components, which resulted in poorer spectral fits.

Results for F05189 are shown in Figure \ref{fig:F05189-1}.
For this AGN the \vcen \ values most clearly distinguish the different kinematic components, unlike for the other the quasars in this study.  A plot of the \vcen \ versus \vsig \ values fit in each spaxel for this quasar (panel a of Figure \ref{fig:F05189-1}) shows multiple distinct groupings of fits separated mostly cleanly by \vcen, with a number of points at high \vsig \ ($>$500 \kms).  
Many of the spectra in individual spaxels are fit with four distinct Gaussian profiles; examples are shown for several spaxels in panel f of Figure \ref{fig:F05189-1}.  

The kinematic components are first sorted by \vcen, with the first component being assigned to the highest \vcen \ value (closest to the stellar redshift of the source) and 
subsequent components assigned to have more blueshifted \vcen \ values, such that the fourth component (c4) is the most blueshifted. For spaxels where fewer than four components are detected, the fits are assigned to kinematic components such that fits with \vcen \ $>$ -200 \kms \ are in the first component, fits with -800 $<$ \vcen \ $<$ -200 \kms \ and \vsig \ $<$500 \kms \ are in the second, fits with -1300 $<$ \vcen \ $<$ -800 \kms \ and \vsig \ $<$500 \kms are in the third, and any remaining fits are are the fourth component (Figure \ref{fig:F05189-1}a).

The first kinematic component has a low ratio of \oiii / \oii \ emission (panel d of Figure \ref{fig:F05189-1}), reflecting that this gas is predominantly tracing star formation in the host galaxy.  Panel b in Figure \ref{fig:F05189-1} shows surface brightness maps of the \oiii \ 5007 \AA \ emission in each kinematic component.  The \oiii \ emission is brighter in the blueshifted components than in the first kinematic component, which is dominated by \oii \ emission from the host galaxy. The first kinematic component has a radial extent of 7 kpc (panel c of \ref{fig:F05189-1}); deconvolving this extent with the seeing reduces it to 6 kpc as listed in Table 2. 

The second kinematic component is 
blueshifted by a median of -540 \kms \ and has a median velocity dispersion of 207 \kms, with a slightly larger physical extent of r$\sim$8 kpc.  The third kinematic component is blueshifted by a median of -937 \kms \ and has a median velocity dispersion of 144 \kms, with a radial extent of r$\sim$6 kpc.  The fourth component has a median blueshift of -1065 \kms \ and a high median velocity dispersion of 634 \kms, with a slightly smaller radial extent of $\sim$5 kpc.  

The \oiii / \oii \ flux ratio is high, $\ge$3, for gas in the second, third, and fourth kinematic components, indicating AGN photoionization of the outflowing gas in each of these components.  Values for the median \vcen, determined across all spaxels, as well as the median \vsig \ and $v_{98}$, and the radial extent (deconvolved with the seeing) for each kinematic component are listed in Table 2. 

Spatial maps of the kinematic properties of each component are shown in panel e of Figure \ref{fig:F05189-1}.  As can be seen, the central velocities differ for each component, and there is spatial coherence to each of the \vcen \ maps, with no strong gradients observed. The \vsig \ maps generally show spatial coherence, with the \vsig \ values being lowest in the first two components. 

The second component is detected across the entire host galaxy, with a small additional extension in the southwest.  The central velocity does not change substantially across the source, indicating that this outflow is detected primarily along the line of sight.  The velocity dispersion is higher than the gas in the first component and is fairly constant spatially across the outflow.

The third component is detected in roughly half of the spaxels, with similar radial extent as the first two components.  The velocity dispersion is higher in the central spaxels, $\sim$350 \kms, and decreases to $\sim$200 \kms \ at larger radius.

The fourth component is detected across the bulk of the source to a radius of $\sim$5 kpc with a high velocity dispersion of $\sim$750 \kms \ in the central spaxels, decreasing somewhat at larger radii. 
In general,  kinematic components with high \vsig \ require high S/N data to both detect and accurately determine the velocity width, as these kinematic components are often seen as low flux asymmetric wings 
in the velocity profiles, which can make it difficult to detect the faster outflowing components at larger radius.  

On the whole the median velocity dispersion is similar for the first three components and is substantially higher for the fourth component, while the median central velocity is much more blueshifted for the third and fourth components than the first and second.

The lack of detection of the third component in many spaxels can be understood by examining the spectra and fits in individual spaxels, as shown in panel f of Figure \ref{fig:F05189-1}.  The lower right spectrum clearly shows four components with different \vcen \ values. In the lower middle spectrum the first component has low flux but is well detected in \oii \ and is real; the other three components are clearly detected in \oiii, with different \vcen \ values.  In the lower left spectrum only three components are detected, though it appears that the component with high \vsig \ (shown in red) could potentially be fit by two separate components if there was sufficient $S/N$.  What is detected as the third and fourth  components in the lower right spectrum here is potentially being fit by a single broader component.  The upper right spectrum is similar, showing three components being fit, while in the upper middle spectrum four components are fit.  The upper left spectrum, at larger radius and with lower S/N, has three components fit. It appears possible that with higher $S/N$ data all four components might be better detected across the source.

We note that in order to derive inferred outflow properties in the next section below we do not utilize the \vsig \ measurements but rather the \vmax \ values, taking the median across all of the spaxels for which that component is detected.  We therefore expect that the possible blending of the third and fourth kinematic components in some spaxels does not substantially impact the inferred maximum velocities, though it does impact the observed flux values.

\subsection{F07559+5058}

\begin{figure*}[t!]
\centering
\includegraphics[width=0.97\textwidth, trim=0.5cm 3cm 1.5cm 0.5cm, clip]{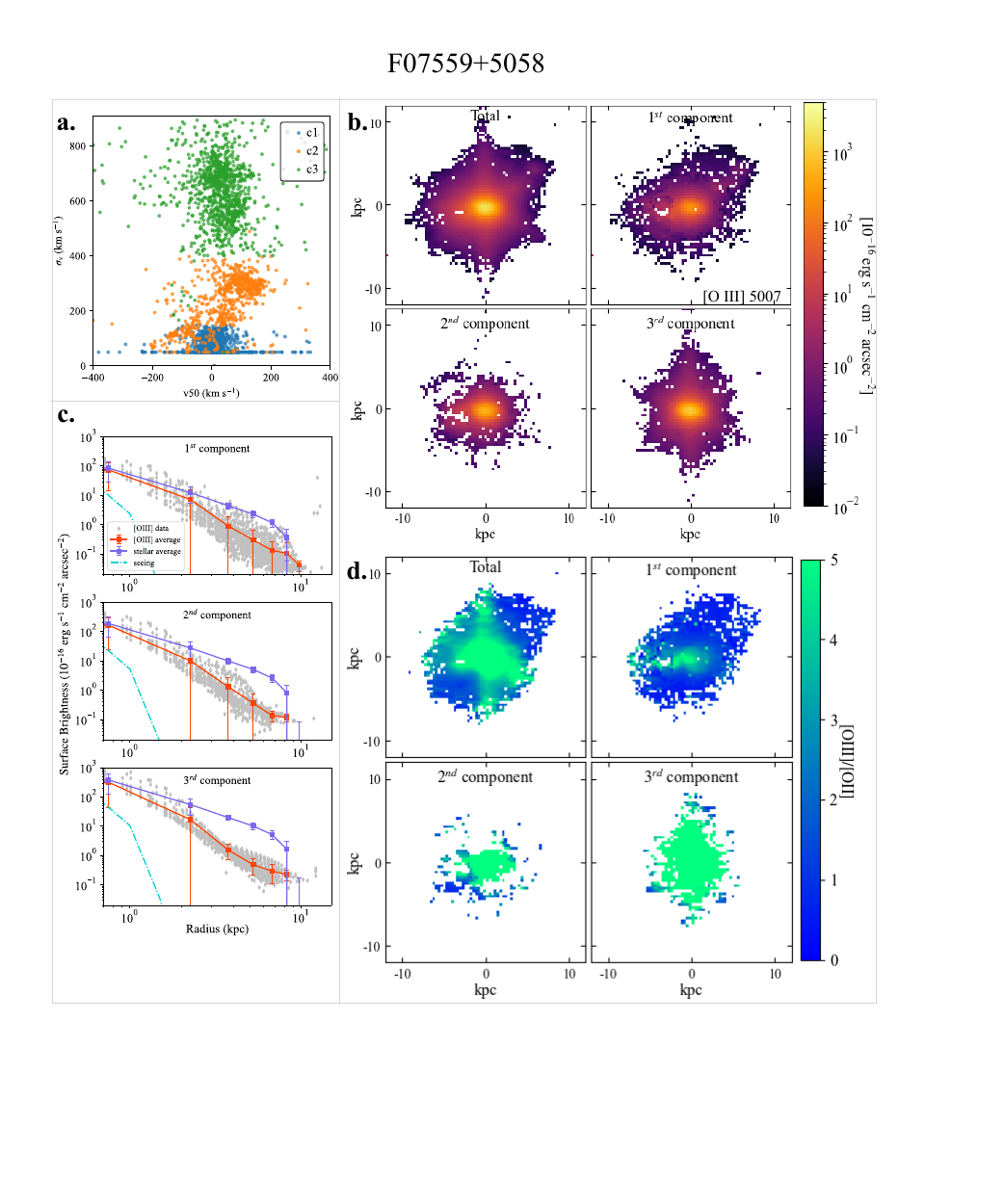}
\caption{Results for F07559+5058: panels show the same information as in Figure 4, shown here for the \oiii \ 5007 \AA \ emission. This quasar has a total of three kinematic components.  The second kinematic component is predominantly redshifted, and the third kinematic component has very high velocity dispersion.}
\label{fig:F07559-1}
\end{figure*}

\begin{figure*}[t!]
\ContinuedFloat
\centering
\includegraphics[width=0.75\textwidth, trim=0.5cm 4.5cm 1.5cm 0.5cm, clip]{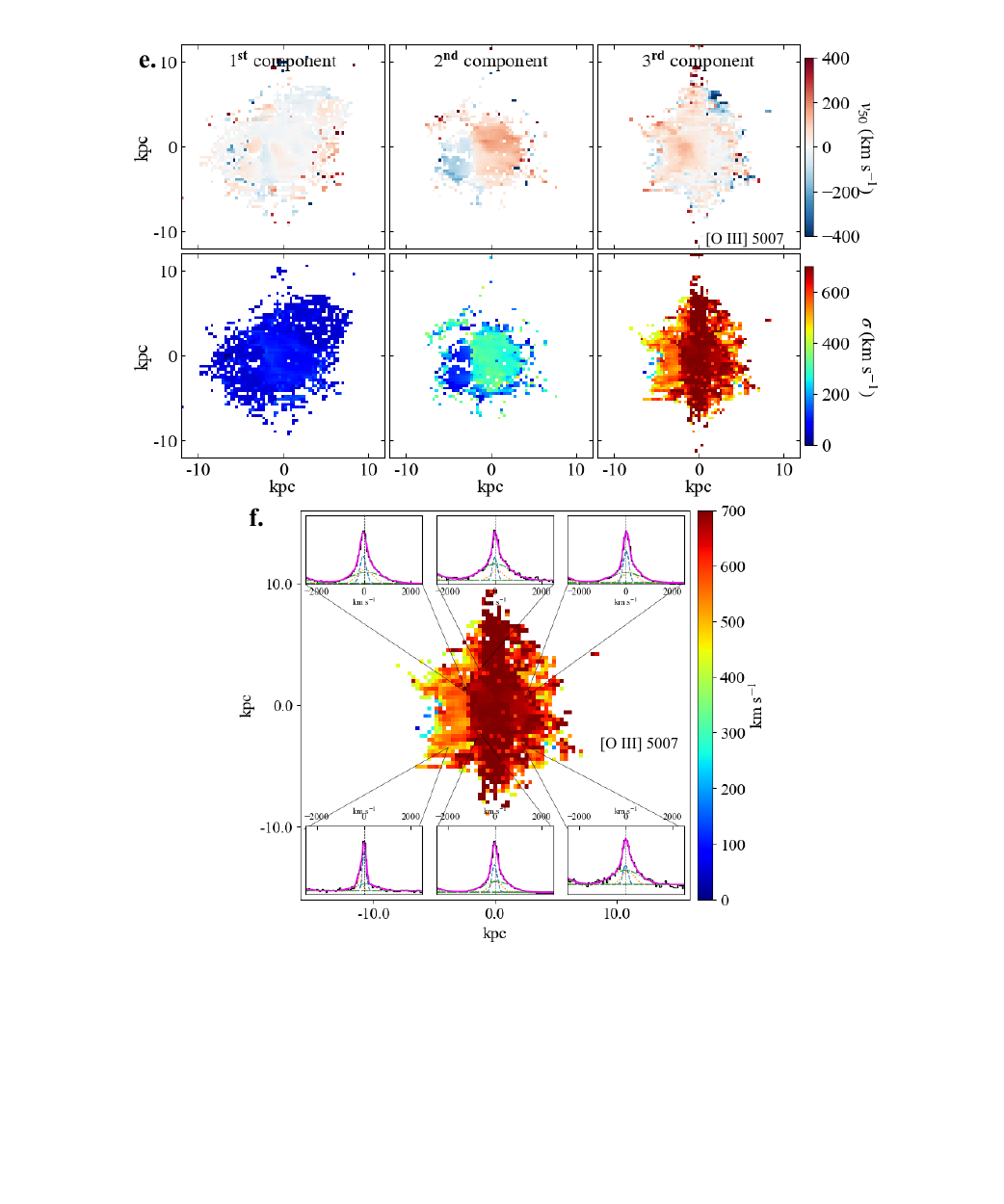}
\caption{Results for F07559+5058 (continued): panels show the same information as in Figure 4, shown here for the \oiii \ 5007 \AA \ emission.}
\label{fig:F07559-2}
\end{figure*}

F07559$+$5058 (hereafter F07559) is an infrared-luminous Type 2 quasar \citep{1987A&AS...70...95D, 2001A&A...374...92V,2008AJ....136.2373R}. \citet{2014MNRAS.440.3202V} find that the nuclear spectrum of F07599 requires three Gaussians to fit \oiii, with two of them very broad (FWHM of 700 and 1700\,\kms). \citet{Fischer18} confirm this with {\it HST} long-slit observations, and measure spatially extended emission in all three components. The maximum projected radial extent of the broadest components is 0.6 kpc. The host galaxy does not show signs of a recent merger in the {\it HST} data. There is a one-sided radio jet of similar size but whose lobe is opposite the nucleus from the brightest \oiii\ emission \citep{2021MNRAS.503.1780J}


The continuum emission map (Figure \ref{fig:cont_line_maps}) for this galaxy is dominated by the quasar, as evidenced by diffraction spikes (particularly along the north-south axis).  The extension seen to the northwest in both the continuum and line emission map is not a diffraction spike; it has narrow \oii, \hb, and \oiii \ emission associated with the host galaxy, which can also be seen extended to the southeast in the line emission map. 

For this source we use IFSFIT to fit  the \textrm{Ne}~\textsc{v}, \oii, \neiii, \hg, \hb, and \oiii \ emission.  We detect three kinematic components in this quasar.  The \vcen-\vsig \ diagram (panel a of Figure \ref{fig:F07559-1}) shows a clustering of points (labeled c1) near the systemic redshift at \vcen$\sim$0 with low \vsig \ values, and a structure (labeled c2) extending to higher \vsig \ ($\sim$300 \kms) that is blueshifted up to $\sim$-200 \kms \ at lower \vsig ($\sim100$ \kms) and redshifted up to $\sim$200 \kms \ at higher \vsig ($\sim$300 \kms), and a plume (labeled c3) to very high \vsig \ ($\sim$800 \kms) near \vcen$\sim0$ \kms. 

To classify each Gaussian profile into one of the three kinematic components, we placed any component with \vsig$<$150 \kms\ into the first component (c1); the second component (c2) fits have $150<$\vsig$<400$ \kms; and any fits with \vsig$>$400 \kms \ are classified as third component (c3).  For example, if the spectrum in a spaxel was fit with only one Gaussian with \vsig$>$400 \kms, it was assigned to the third component.  

The \oiii / \oii \ line ratios for each kinematic component (panel d of Figure \ref{fig:F07559-1}) show that the first component has low \oiii / \oii \ beyond the nucleus, consistent with the emission being dominated by host galaxy light, while the second and third components have high \oiii / \oii, consistent with AGN photoionization. 

The  spatial maps of the kinematics in each component (panel e of Figure \ref{fig:F07559-1}) show that the first component, dominated by emission from the host galaxy, has relatively narrow \vsig (median 64 \kms) and only small variations in \vcen\ ($<$50\,\kms\ peak-to-peak). There is no clear rotational signature, perhaps indicating a face-on system. The spatial extent of this first component emission is $\sim$10 kpc. 

The second kinematic component is redshifted in the center of the galaxy by $\sim$100 \kms \ and to the west up to $\sim150 \kms$, while the gas to the east is blueshifted up to $\sim-150 \kms$.  The emission in this component has a median \vsig $= 262$ \kms \ and a radial extent of $\sim$8 kpc. This outflow appears to be more across rather than along the line of sight, as we are able to detect both the front and back of the outflow as blueshifted and redshifted emission. The directionality of the blue- and red-shifted sides, as well as the roughly E-W elongation of the morphology of this component, are consistent with the E-W alignment at radii $<$1\,kpc of both \oiii\ \citep{Fischer18} and radio emission \citep{2021MNRAS.503.1780J}. The sign of the Doppler shift is also consistent with the intermediate-$\sigma$ component observed by \citet{Fischer18}. However, we observe it extending up to scales of $\sim$7~kpc.

The third kinematic component, in contrast, appears redshifted to the east by $\sim$100 \kms\ and blueshifted to the west by $\sim$-50 \kms, and has a very high velocity dispersion (median \vsig=654 \kms). However, this component appears to be dominated by nuclear emission, as seen by the strong, unremoved diffraction spikes. The radial extent is thus overestimated from the azimuthal average ($\sim$8 kpc, panel c), and we assign an upper limit of 3\,kpc. We take the overall maximum radius for the outflowing gas in this source from the intermediate-velocity component (c2).

\subsection{F07598+2601}

\begin{figure*}[t!]
\centering
\includegraphics[width=0.97\textwidth,trim=0.5cm 3cm 1.5cm 0.5cm, clip]{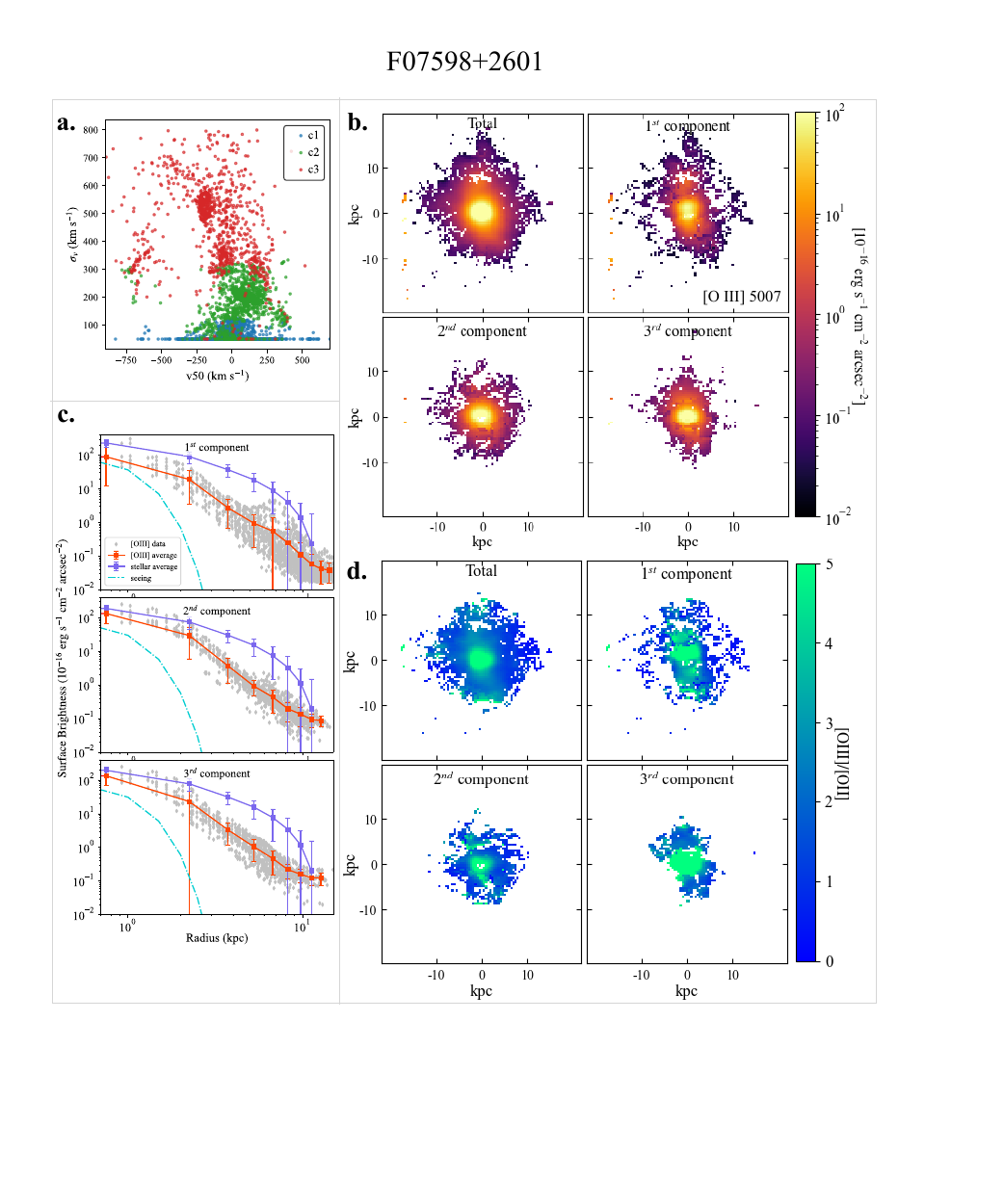}
\caption{Results for F07598+2601: panels show the same information as in Figure 4, here for the \oiii \ 5007 \AA \ emission in this source.  This quasar has a total of three kinematic components. }
\label{fig:F07598-1}
\end{figure*}

\begin{figure*}[t!]
\ContinuedFloat
\centering
\includegraphics[width=0.75\textwidth,trim=0.5cm 3.2cm 1.5cm 0.5cm, clip]{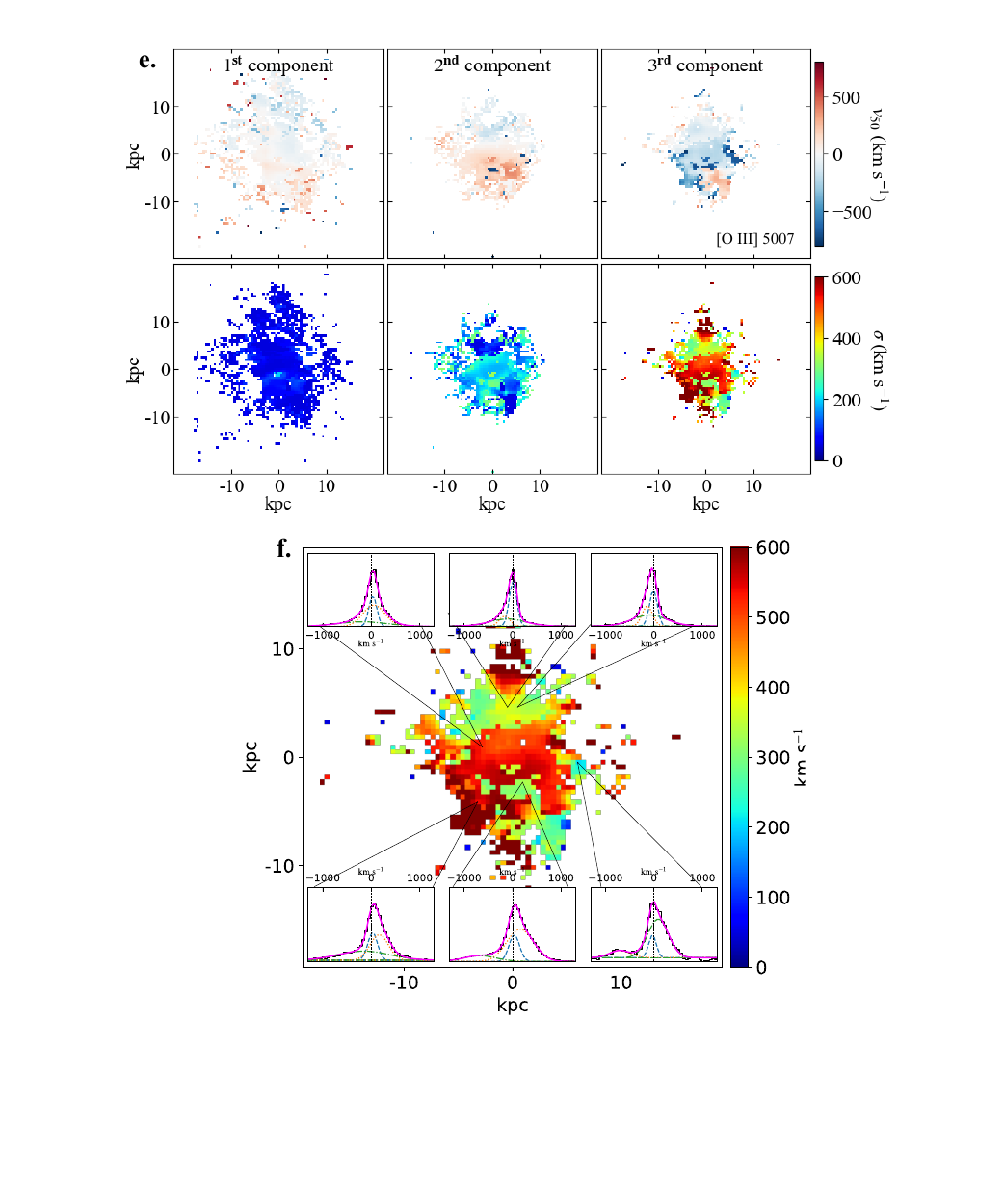}
\caption{Results for F07598+2601 (continued): panels show the same information as in Figure 4, shown here for the \oiii \ 5007 \AA \ emisssion in this quasar.}
\label{fig:F07598-2}
\end{figure*}

Similar to F07559, F07598+2601 (hereafter F07598) is an infrared-luminous Type 2 quasar that was first observed spectroscopically with SDSS \citep{1987A&AS...70...95D,2008AJ....136.2373R}. Just like F07559, it was observed by \citet{2014MNRAS.440.3202V} and \citet{Fischer18} to have three components (narrow, intermediate, and broad) with maximum projected extent of 0.4\,kpc. In this case the \oiii\ elongation is N-S and the broad component is blueshifted by 700\,\kms\ in the nucleus.

This host galaxy is a late-stage major merger, and there are tidal tails observed in the KCWI continuum emission maps (Figure \ref{fig:cont_line_maps}).  For this source we use IFSFIT to fit  the \textrm{Ne}~\textsc{v}, \oii, \neiii, \hg, \hb, and \oiii \ emission.  To increase the S/N at larger radius we ran IFSFIT on a 2x2 boxcar smoothed mosaic. 

We detect three kinematic components in this source. The \vcen-\vsig \ diagram (panel a of Figure \ref{fig:F07598-1}) shows an interesting structure, with both blueshifted and redshifted \vcen \ values up to \vsig $\sim$400 \kms, with a plume of predominantly blueshifted fits extending to higher values of \vsig $\sim$800 \kms.

The kinematic components are first sorted by \vsig, with the first component being assigned to the lowest \vsig \ value, the second component is assigned to the next larger \vsig \ value, and the third component assigned to the highest \vsig \ value.  For spaxels where only one or two components are detected, the fits are assigned to kinematic components based on \vsig \ such that  fits with \vsig $<$ 120 \kms \ are assigned to the first component, fits with 120 $<$ \vsig $<$ 320 \kms \ are assigned to the second component, and fits with \vsig $>$ 320 \kms \ are assigned to the third component. 

The first component is dominated by emission from the AGN and the host galaxy extending to r$\sim$14 kpc; the spatially-extended emission in this component shows clear rotation, blueshifted by $\sim$100 \kms in the northwest (see panel e of Figure \ref{fig:F07598-1}) and redshifted by $\sim100$ \kms \ in the southeast. The \oiii / \oii \ ratio for this kinematic component is high near the center of the  galaxy, reflecting AGN photoionization, and falls to lower values more typical of star formation at larger radius. 

There are two additional kinematically distinct components in this source, one predominantly redshifted and one blueshifted.  The second kinematic component is redshifted by $\sim$250 \kms \ across the bulk of the spatial extent (top middle of panel e), including in the south and the nucleus, and it is  blueshifted up to $\sim$150 \kms \ in the north.  This kinematic component has a typical velocity dispersion of \vsig$\sim$300 \kms \ and a radial extent of $\sim$11 kpc.  

The third kinematic component is a faster outflow which is blueshifted by $\sim$200 \kms \ across the source (top right of panel e) and has a velocity dispersion of \vsig$\sim$500 \kms \ and a radial extent of $\sim$11 kpc.  (The small redshifted region in the southwest may more realistically be associated with the second kinematic component; in this region the spectra were typically fit with two components with low \vsig \ and a third component with higher \vsig).
In this source the two outflowing components are observed to extend to larger radii than the emission from the stars (panel c).

While for the rest of the quasars in this sample the presence of distinct multiple kinematic components beyond the first component (associated with the host galaxy) appears to correspond to separate outflows, it is not clear for F07598 if the second and third kinematic components are distinct outflows or the front (blueshifted) and back (redshifted) sides of a single outflow. The overall flux distribution of \oiii \ in this source is clearly oriented N-S, which is an extension to $\sim$10 times larger radius what is seen in \citet{Fischer18}. There is evidence in the \oiii / \oii \ ratio for AGN photoionization along the N-S axis in all three kinematic components seen here (panel d). The velocity gradient in the second component is consistent with a biconical outflow along the ionization cone. 

The third component could be interpreted as being related, in that the outflow cone could be oriented partly in the plane of the sky, and the very high \vsig \ values in the third component could be related to  the base of the outflow where there is potentially a combination of interaction with the host disk (oriented along the line of highest dispersion) as well as overlap of the two outflow cones in the center of the source. These kinematic maps are somewhat similar to those of the outflow observed in a Type 1 quasar at redshift $z=0.435$ of \citet{Shen23}, where redshifted and blueshifted emission are observed on opposite sides of the source to similar scales as seen here, with a distinct region of high velocity dispersion running perpendicular to the biconical outflow in the center.
The KCWI results for F07598 shown here do not present as clean of an interpretation as the source in \citet{Shen23}, in part because the third component has a substantial bulk blueshift (seen in the \vcen \ map in the upper right of panel e) across most of the source, and at large radius two blueshifted components are detected in the north and two redshifted components are detected in the south.  The physical picture in this quasar may be more complex, with contributions from both a biconical outflow and two distinct outflows.  For consistency with the rest of the sample we report the radial extent and kinematics of each kinematic component for this source in Table 2.

\subsection{J1010+0612}

\begin{figure*}[t!]
\centering
\includegraphics[width=0.97\textwidth, trim=0.5cm 3cm 1.5cm 0.5cm, clip]{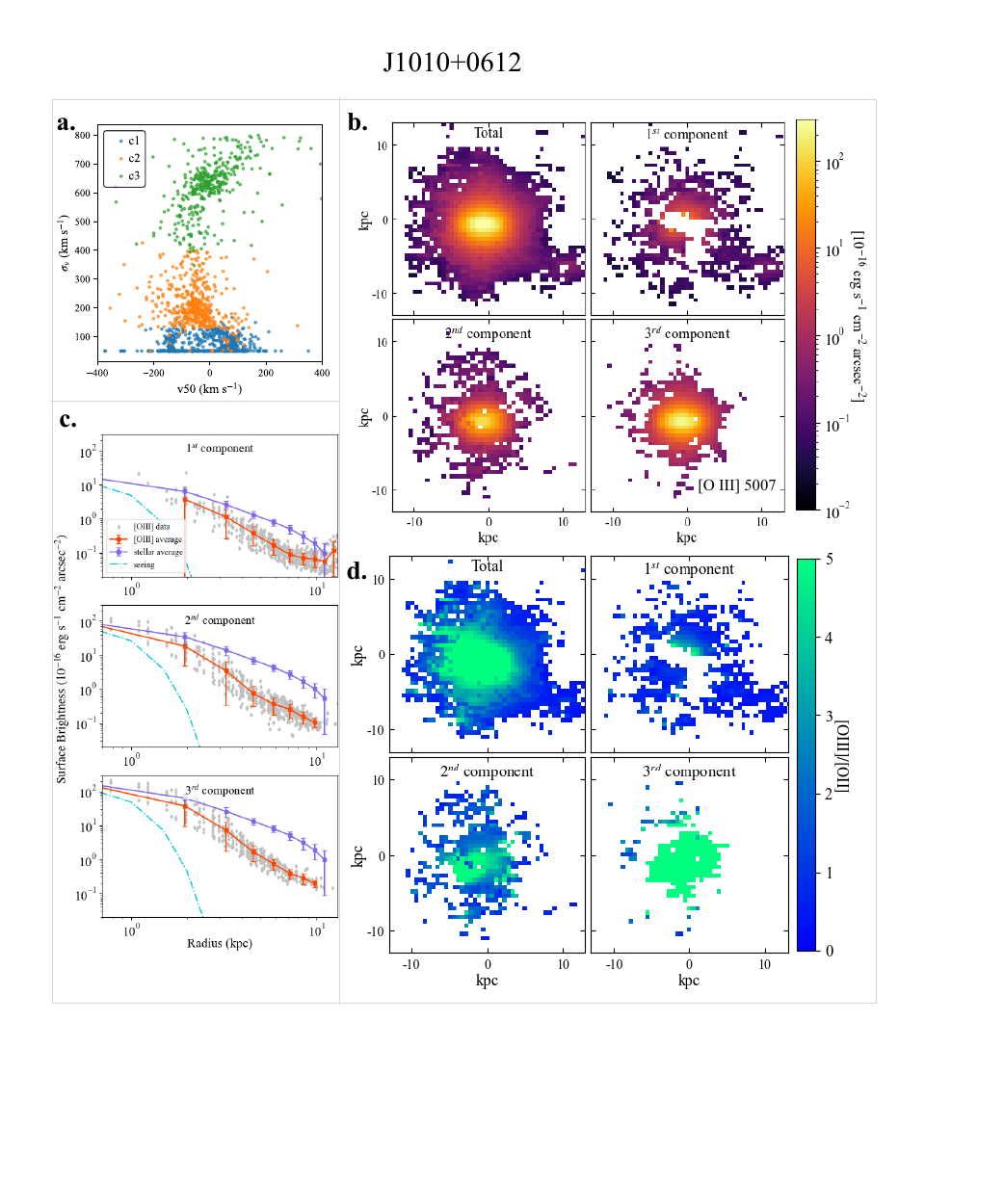}
\caption{Results for J1010+0612: panels show the same information as in Figure 4.  This quasar has a total of three kinematic components. }
\label{fig:J1010-1}
\end{figure*}

\begin{figure*}[t!]
\ContinuedFloat
\centering
\includegraphics[width=0.75\textwidth, trim=0.5cm 4cm 1.5cm 0.5cm, clip]{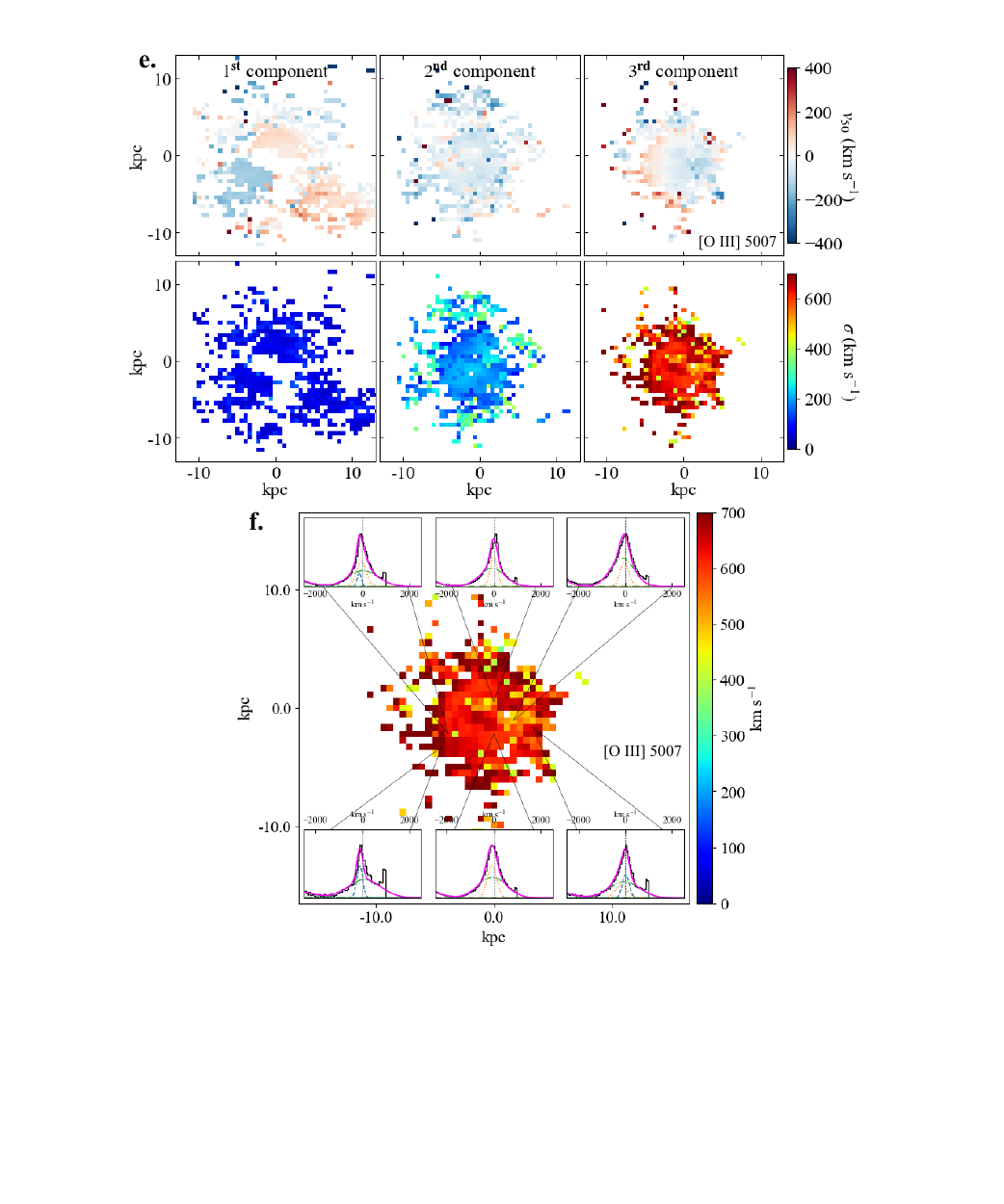}
\caption{Results for J1010+0612 (continued): panels show the same information as in Figure 4.}
\label{fig:J1010-2}
\end{figure*}

J1010+0612 (hereafter J1010) is an optically-selected Type 2 quasar \citep{2001A&A...374...92V,2008AJ....136.2373R}. It was observed by \citet{2014MNRAS.440.3202V}, who measured three kinematic components in \oiii\ in the nucleus. All three are broad and modestly blueshifted, with FWHM of 400, 1200 and 3400\,\kms\ and central velocities of $-$11, $-$66 and $-$250\,\kms, respectively. IFS observations paint a consistent picture of a mostly circular \oiii\ nebula extending to radii of 3\arcsec, or 6.5~kpc \citep{Harrison14,2019MNRAS.485.2710J,Ulivi24,Bianchin26}. There is a velocity gradient NW to SE (red to blue) that is consistent with the rotation of its stars \citep{Ulivi24}. The inner nebula is broad ($W_{80}\sim1100$ \kms) and blueshifted by 100~\kms\ \citep{Harrison14}, consistent with the nuclear emission. J1010 is also a rare example of a Type 2 quasar with silicates in emission in the mid-IR, suggesting that the torus may be outflowing and that the obscuring screen is being blown away \citep{2026A&A...706A.100R}. The galaxy is also involved in a minor merger with a small companion to the SW (12.4~kpc distant) and a long tidal tail in the same direction \citep{2019AJ....157..168D}.

The KCWI data for this source covers the \oiii \ 4959 and 5007 \AA \ lines, though the 5007 \AA \ line is at the edge of the detector, such that the reddest portion of the line at velocities $>$1000 \kms \ is not covered (this can be seen in panel f of Figure \ref{fig:J1010-1}, where the spectral data abruptly ends in the insets showing the spectra and Gaussian fits to the \oiii \ 5007 \AA line). We use IFSFIT to fit \oii, \neiii, \sii, \hg, \hb, and \oiii \ emission in this source and find three kinematic components.

Consistent with previous work, we find that the host galaxy of J1010 has a circular continuum morphology (Figure \ref{fig:cont_line_maps}) and a diffuse low surface brightness extension of stellar continuum and line emission extending towards the east.  We detect the interacting secondary galaxy, which like the primary contains \oii \ emission (Figure~\ref{fig:cont_line_maps}).

The \vcen-\vsig \ diagram (panel a of Figures \ref{fig:J1010-1}) shows three regions of fits: those with low \vsig \ that are  predominantly redshifted, fits that are blueshifted by $\sim$-50 \kms \ with \vsig$\sim200$ \kms, and a plume of fits to higher \vsig \ (up to 800 \kms) that extend from roughly near the systemic velocity to more redshifted at higher velocity dispersion. When assigning Gaussian profiles to kinematic components, we first sort the fits in a given spaxel such that the higher \vsig fits are assigned to higher components.  If less than three Gaussian profiles are fit, fits are assigned to the first component (c1) if \vsig$<130$ \kms; to the second (c2) if $130$ \kms \ $<$ \vsig $<400$ \kms; and to the third (c3) if \vsig $>400$ \kms.

The first kinematic component is dominated by emission from the host galaxy, extending to $\sim$11 kpc with low \oiii / \oii \ ratios (panel d).  Many spaxels near the central AGN do not contain $c1$, which means a low-$\sigma$ component. In examining the fits in these spaxels it is seen that the observed flux in \oiii \ is dominated by the outflowing gas and little gas near the systemic redshift is detected, consistent with the lack of a narrow component in nuclear spectra \citep{2014MNRAS.440.3202V}.  Also consistent with previous work, the first component shows rotation in the spatial kinematic maps, with gas to the southeast blueshifted by $\sim-130$ \kms \ and gas in the north redshifted by $\sim70$ \kms  \citep{Ulivi24}. The gas in the southwest, including the gas in the nearby galaxy, is redshifted by $\sim$100 \kms.

The second kinematic component is blueshifted by $\sim-70$ \kms \ and has a median velocity dispersion of \vsig$=194$\,\kms, extending to a radius of $\sim$10 kpc.  The third component is redshifted towards the east up to $\sim$100 \kms, slightly blueshifted in the center of the source with \vcen \ $\sim-20$ \kms, and blueshifted to the west up to $\sim-100$ \kms. This component, c3, has a high velocity dispersion with a median value of \vsig$=630$\,\kms \ and a radial extent of $\sim$10 kpc.

In this quasar the second and third kinematic components appear to be tracing two distinct outflows with differing kinematics, but to higher radii than previously known. The third component also has a gradient that could indicate a flow along the galaxy's minor axis; i.e., perpendicular to the rotation axis. However, the orientations of these axes are not well-constrained.

\subsection{J2133+0712}

\begin{figure*}[t!]
\centering
\includegraphics[width=0.97\textwidth, trim=0.5cm 3cm 1.5cm 0.5cm, clip]{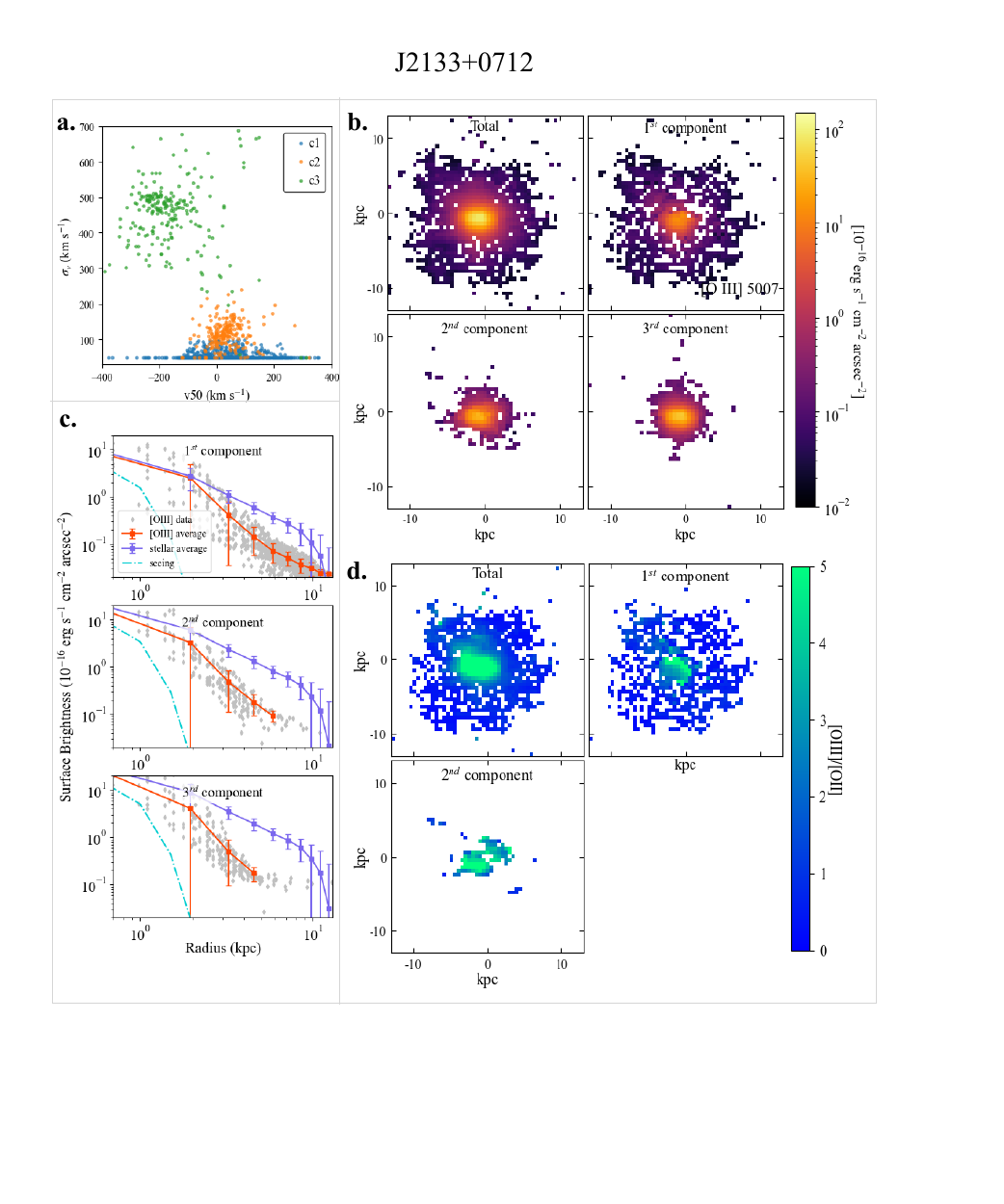}
\caption{Results for J2133+0712: panels show the same information as in Figure 4.  This quasar has a total of three kinematic components. }
\label{fig:J2133-1}
\end{figure*}

\begin{figure*}[t!]
\ContinuedFloat
\centering
\includegraphics[width=0.75\textwidth, trim=0.5cm 4cm 1.5cm 0.5cm, clip]{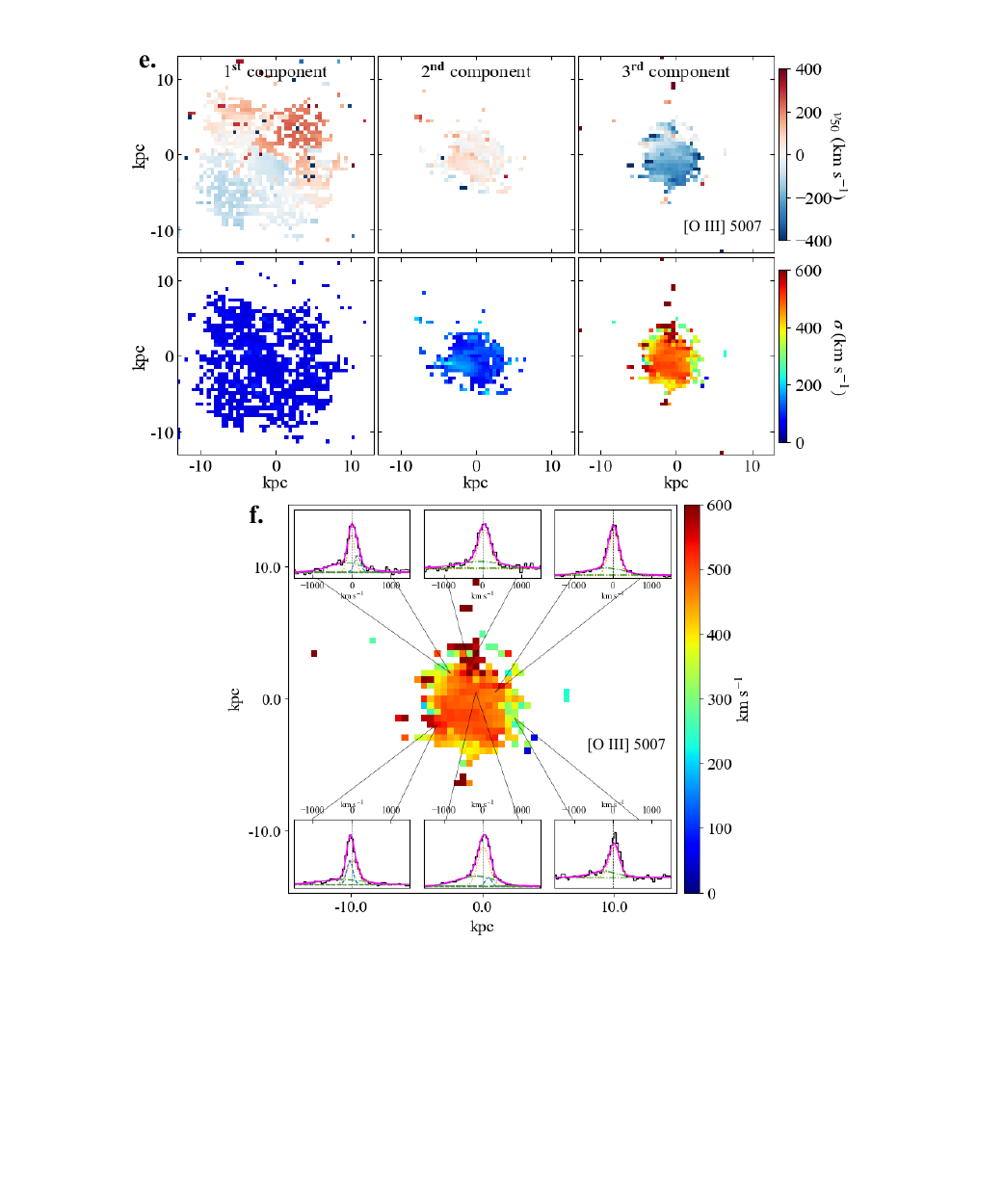}
\caption{Results for J2133+0712 (continued): panels show the same information as in Figure 4. The second kinematic component in this quasar is redshifted, while the third kinematic component is blueshifted and has higher velocity dispersion.}
\label{fig:J2133-2}
\end{figure*}

J2133+0712 (hereafter J2133) is a star-forming spiral galaxy with a Type 2 quasar \citep{2006AJ....132...27B,Sun17}. It has a radio excess \citep{2003AJ....126.2237D}, with $L_{1.4GHz} = 10^{24.5}$\,W\,Hz$^{-1}$ \citep{2024ApJ...973...48C}, and is classified by \citet{2012MNRAS.421.1569B} as a compact, high-excitation radio galaxy. It exhibits a narrow, slightly-redshifted \ion{H}{1} absorber \citep{2024ApJ...973...48C}.

The outflow in J2133 was detected in \citet{Kang18} using Gemini GMOS IFS data; they find blueshifted \oiii \ emission with a flux-weighted velocity centroid of -104 \kms \ and \vsig \ of 420 \kms, with a radial extent of 1.2 kpc.  In Magellan long-slit spectroscopy, the outflow in this source is spatially unresolved \citep{Sun17}.

The KCWI continuum map of J2133 presented here, which is dominated by light from the host galaxy and the AGN, shows some spiral structure, particularly to the southeast, where knots of star formation can be seen in the \oii \ line emission map as well.  For this source we use IFSFIT to fit \nev, \oii, \neiii, \hg, \hb, and \oiii \ emission and find three kinematic components.  

The \vcen-\vsig \ diagram (panel a in Figure \ref{fig:J2133-1}) shows fits with low \vsig \ extending across a wide range of \vcen \ values centered near the systemtic redshift, along with a region of fits with low \vcen \ and somewhat higher \vsig $\sim120$ \kms. A third grouping of blueshifted fits is seen 
with \vcen$\sim-200$ \kms \ and  high velocity dispersion of  \vsig$\sim$500 \kms.

When assigning Gaussian profiles to kinematic components, we first sort the fits in a given spaxel such that the higher \vsig fits are assigned to higher components.  If less than three Gaussian profiles are fit, fits are assigned to the first component (c1) if \vsig$<100$ \kms \ and to the second kinematic component (c2) if $100$ \kms \ $<$ \vsig $<220$ \kms \ and to the third kinematic component (c3) if \vsig $>220$ \kms.

The first kinematic component is dominated by emission from the host galaxy, extending to $\sim$10 kpc with low \oiii / \oii \ ratios (panel d); this ionization ratio is higher in the center of the galaxy for the first component. This component is well fit by relatively strong \oii \ emission; the \oiii \ emission at larger radius in the host galaxy is often observed at low $S/N$, resulting in missing spaxels in the maps of \oiii \ emission (panels b and d). There is clear rotation seen in the spatial kinematic map of the first component (panel e of Fig.\ref{fig:J2133-1}), with blueshifted emission to the southeast with \vcen$\sim-100$ \kms \ and redshifted emission to the northwest with \vcen$\sim$200 \kms.

The second kinematic component has a higher velocity dispersion with a median \vsig$=114$ \kms \ with a radial extent of $\sim$6 kpc.  The \vcen \ values for this component are near systemic though somewhat redshifted on average; the gas is redshifted by $\sim50$ \kms \ in the southeast (where the first kinematic component is blueshifted) and  blueshifted  by $\sim30$ \kms \ in the northwest (where the first kinematic component is redshifted). This outflow appears to be observed with some projection across the line of sight. The redshifted absorber seen in \ion{H}{1} could be associated with this component \citep{2024ApJ...973...48C}.

The third kinematic component is blueshifted across the spatial extent of the outflow, with an extent of $\sim$4 kpc and a median \vcen$=-181$ \kms \ and a high velocity dispersion with a median \vsig$=471$ \kms.  This outflow is seen primarily along the line of sight.


\section{Derived Outflow Mass, Energy, and Momentum}
\label{sec:derived-properties}

\input{derived_properties_table.tex} 

The mass and energy carried by the observed outflows presented here are important for understanding the impact these AGN-driven outflows have on their host galaxies.
Using \oiii \ as a tracer of the AGN-driven outflows (\hb \ in the case of I~Zw~1 and F01004), we use the spatial and kinematic features of the second, third, and---in the case of F05189---the fourth velocity components to determine the radial extent and maximum velocity of each component. Combined with gas mass estimates derived from the \hb \ luminosity in these components, we obtain first-order estimates of the mass, energy, and momentum of the ionized gas in these outflows. 

First, we estimate the color excess $E(B-V)_\mathrm{gas}$ from the Balmer decrement, assuming case B, the \citet{Cardelli89} extinction law, and $R_V = 3.1$. As measurements of the Balmer decrement in each spaxel are highly uncertain due to the relatively low flux of the \hg \ emission line, we use spatially-integrated spectra with a radius of 8 spaxels to create high $S/N$ spectra for each AGN from which to measure the Balmer decrement. 
We use the total flux in the \hb \ and \hg \ emission lines, not divided into separate kinematic components. Taking the intrinsic line ratio of H$\gamma$/H$\beta$ = 0.47, we use the observed line ratio to calculate an average reddening in the gas due to dust. The estimates of $E(B-V)$ for each source are listed in Table 4.  Errors on the $E(B-V)$ values are calculated from the errors on the \hb \ and \hg \ fluxes in the spatially-integrated spectra.
The $E(B-V)$ values derived in the central spaxels of the mosaics typically have similar values to those determined from the spatially-integrated spectra, except in two cases: for F07598 the value of $E(B-V)$ in the central spaxels is 0.30, while the value from the spatially-integrated spectrum is 0.67, and for F07559 the value in the central spaxels is 0.06, while the value from the spatially-integrated spectrum is 0.97.  In F07559 in particular the Balmer decrement is substantially higher at larger radius compared to smaller radius; this strong radial dependence is not observed in the other AGN.

We then apply the dust correction to the summed total fluxes of \hb \ in the outflow components (c2, c3, c4).  We do not apply an additional Galactic extinction correction to the fluxes, given that it is generally subdominant to the dust exinction in the AGN host galaxy. The fluxes are then converted into luminosities,  listed in Table 4, along with the observed and dust-corrected \hb \ luminosities.  For F07598, the two components could be emission from the front and back sides of the same biconical outflow, but we report these two components separately in Table 4 for consistency with the rest of the sample .

Next we estimate the mass of the ionized outflowing gas by calculating the mass of recombining hydrogen atoms, assuming purely photoionized gas with Case B recombination
with an intrinsic line ratio of $H\alpha / H\beta= 2.9$ and an
electron temperature of $T = 10^4\,K$, following \citet{Osterbrock06}  and \citet{Nesvadba17}.
The mass is given by 
\begin{equation}
    M_{\text{out}} = 9.5\times 10^8 \ \text{M}_{\odot} \left(\frac{L_{\text{H$\beta$,out}}} {10^{43} \ \text{erg s}^{-1}}\right) \left(\frac{n_e}{100 \ \text{cm}^{-3}}\right)^{-1}
    \label{eqn:Mion_from_hbeta_lum}
\end{equation}
where $L_{\text{H$\beta$,out}}$ is the outflow \hb \ luminosity and $n_e$ is the electron density of the ionized gas. Table 4 lists the calculated mass of the ionized gas in each outflow component, assuming $n_e = 100 \, \text{cm}^{-3}$ as a fiducial value; the uncertainty on this value is discussed below. 

The energy and momentum of the ionized outflowing gas, as well as the respective outflow rates with time, are calculated using the radial extent and velocity measurements of each kinematic component of the outflow. The kinetic energy of the ionized outflow is given by 
\begin{equation}
    E_{\text{kin}} = \frac{1}{2}M_{\text{out}}v_{\text{out}}^2
    \label{eqn:KE_outflow}
\end{equation}
where we use $v_{\text{out}} = v_\mathrm{max,med}$, the median across all spaxels of the maximum velocity.

The mass outflow rate is obtained by
\begin{equation}
    \dot{M}_{\text{out}} = M_{\text{out}} \times \frac{v_{\text{out}}}{R_{\text{out}}}
    \label{eqn:mass_outflow_rate}
\end{equation}
where $R_{\text{out}}$ is the radial extent of the outflow, deconvolved with the seeing, as analyzed in the previous section and listed in Table 2.  For the third component of F07559, where we measure only an upper limit of 3 kpc for the radial extent, we set $R_{\text{out}} = 0.7$\,kpc \citep{Fischer18}.

An overall factor of $1-3$ is sometimes applied to the above equation as a stand-in for the geometry of the outflow \citep{Harrison18}.
A spherical outflow implies an additional factor of 3, while an outflow covering 1/3 of the entire sphere provides a factor of 1. 
In this study, we use a fiducial value of 1. Using a different factor would simply scale the overall mass and energy outflow rates.

The kinetic energy outflow rate and momentum flux are then given by
\begin{equation}
    \dot{E}_{\text{kin}} = \frac{1}{2}\dot{M}_{\text{out}}v_{\text{out}}^2
    \label{eqn:KE_outflow_rate}
\end{equation}
and 
\begin{equation}
    \dot{P}_{\text{out}} = \dot{M}_{\text{out}} v_{\text{out}}
    \label{eqn:momentum_outflow_rate}
\end{equation}

Table 4 summarizes our calculated values of the mass, mass outflow rate, energy, energy outflow rate, and the momentum flux for each outflow in each quasar.  The mass of the ionized outflow in the third component in F05189 is likely underestimated due to this outflow not being detected separately in many of the spaxels in this source, and therefore the flux is underestimated.  This leads to the derived outflow properties (mass, energy, momentum) being lower limits for this outflow. In contrast, the values derived for the fourth component in this AGN may be upper limits, due to an additional contribution from the third outflow being assigned to the single Gaussian detected in some spaxels in this source.

We note that \citet{Rupke05} advocate that the kinetic energy and kinetic energy outflow rate equations include a term for the `turbulent' kinetic energy, $3/2 \ \sigma^2$, in addition to the $1/2 \ v_{out}^2$ bulk kinetic energy term in equations 2 and 4 above.  We do not include a turbulent energy term in the kinetic energy values derived here, as the outflow velocity used in these equations in this paper is estimated using $v_{98}$, which includes \vsig.  Similarly, \citet{Ulivi24} use an outflow velocity of $v_{out} = v_{50} + 2 \ \sigma$ and do not include the turbulent term in their outflow energy measurements, as is done here.  If instead for the kinetic energy and kinetic energy rates we use \vcen \ as the bulk velocity and add the turbulent kinetic energy term, the energy estimates listed in Table 4 decrease by 33\%.  

In the above calculations we assume $n_e = 100 \, \text{cm}^{-3}$ to allow for comparison with values reported in the literature \citep[e.g.,][]{Rupke17}. The outflow mass scales inversely with $n_e$ such that. If, for example, $n_e = 500 \, \text{cm}^{-3}$, the mass, energy, and momentum values, as well as their outflow rates, all decrease by a factor of 5. 

We emphasize that the values listed in Table 4 are highly uncertain and are presented here in order to facilitate both comparison with the literature as well as to make relative comparisons between the multiple outflows observed within a given AGN.  The radial extent of the outflows that is observed depends on the S/N in the data and so could be considered a lower limit. 
More importantly, the election density in the outflowing gas is highly uncertain; this is the dominant source of error in determining these physical parameters.

The value of $n_e$ is difficult to measure as it typically requires detection of the outflow in a density-sensitive line such as the \sii \ doublet, which is not covered in the KCWI data presented here. Measurements of $n_e$ in AGN-driven outflows in the literature span a wide range of values and often have large uncertainties.  \citet{Harrison14} find $n_e \approx 200-1000~\mathrm{cm}^{-3}$ in a sample of $z < 0.2$ luminous AGNs, \citet{kak18} find spatially-resolved electron densities of $\approx 50 - 2000~\mathrm{cm}^{-3}$ in $z < 0.02$ radio AGNs, and \citet{Rupke17} find spatially-averaged electron densities of $50 - 400~\mathrm{cm}^{-3}$, with a median of $150~\mathrm{cm}^{-3}$ in a sample of $z < 0.3$ quasars.  \citet{min19} present high resolution maps of the electron density in several nearby AGNs with outflows  and obtain a range of densities from 50 to 1000 $\mathrm{cm}^{-3}$, with a median of $250~\mathrm{cm}^{-3}$.  Measurements using auroral and transauroral emission line ratios yield higher densities of $10^{3-4.8}~\mathrm{cm}^{-3}$ \citep{hol11, ros18, Davies20}, such that 
\citet{Holden26} argue that the electron densities derived using \sii \ should be scaled up by 0.75 dex.

An additional consideration is that spatially-resolved electron density maps in \citet{Rupke17}, \citet{kak18} and \citet{min19} show that regions of elevated electron densities ($\gtrsim 1000~\mathrm{cm}^{-3}$) are concentrated in small spatial regions at radii $< 0.5$ kpc, while the electron density drops quickly with radius, reaching $< 50 ~\mathrm{cm}^{-3}$ beyond 1 kpc from the AGN. Therefore, higher values of $n_e$ near $\sim 1000 ~\mathrm{cm}^{-3}$ reported from spatially-averaged analyses can be biased towards high density regions which dominate the total flux \citet{min19}. This can potentially lead to overestimation of the electron density over the large spatial volumes traced by the outflows detected in this paper.
While most of the high density gas in the outflows are concentrated in the most luminous compact region, substantial mass can reside in lower density regions with a much higher volume filling factor. 

The results of these studies, taken together, imply that there is at least an order of magnitude uncertainty in the value of the gas density in the outflows, such that the derived mass, energy, and momentum have at least as much uncertainty.  In addition, we report only on the ionized phase of the outflows in these AGN; the properties summed across all gas phases will be higher, substantially so in some cases \citep{Rupke17,2021MNRAS.505.5753F}. In the next section we proceed with comparing the relative values of the derived outflow properties derived here, comparing the multiple outflows detected for a given AGN.

\section{Discussion}
\label{sec:discuss}

\begin{figure*}
    \centering
    \includegraphics[width=0.6\textwidth]{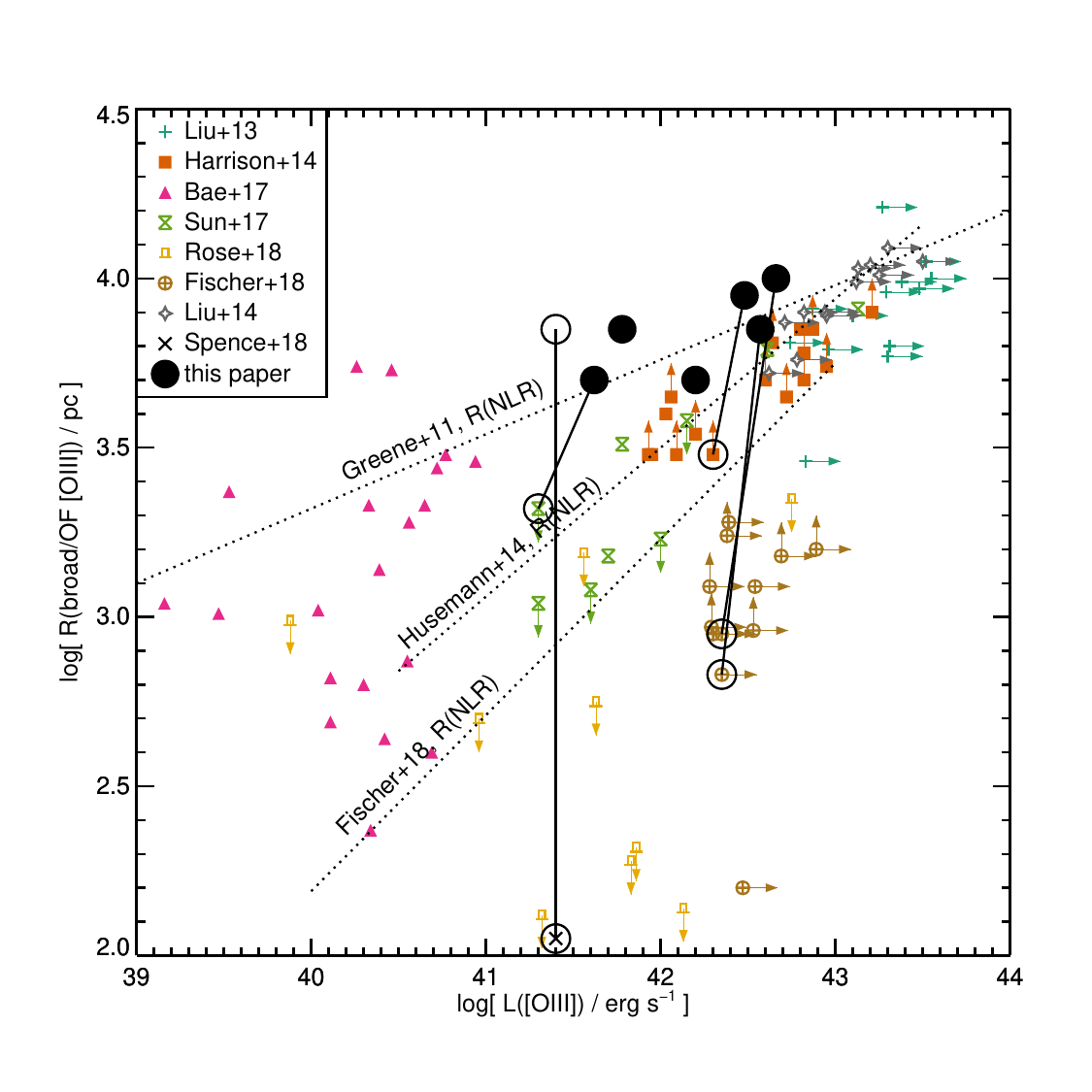}
    \caption{Maximum radial size of the region showing broad and/or blueshifted \oiii\ emission versus total observed (extincted) \oiii\ luminosity. The legend shows the samples to which we compare \citep{2013MNRAS.430.2327L,Harrison14,2014MNRAS.442.1303L,Bae17,Sun17,ros18,Fischer18}, including the measurement of F01004 from \citet{2018MNRAS.478.2438S}. Three determinations of the NLR \oiii\ radius-luminosity relations are shown as dashed lines \citep{2011ApJ...732....9G,2014MNRAS.443..755H,Fischer18}. Our KCWI measurements of both radius and $L_\mathrm{[OIII]}$ are shown as filled solid circles except for F01004, for which we use an open circle to represent the point determined by our radial measurement and the luminosity from   \citet{2018MNRAS.478.2438S}. Where available, previous measurements of galaxies in our sample are circled and connected to our KCWI measurements by solid lines.}
    \label{ro3-vs-lo3}
\end{figure*}

\subsection{Outflow Sizes and Physical Properties} \label{subsec:ofprop}

In the sample of seven nearby quasars studied here using KCWI, we identify a total of 13 separate outflowing velocity components. The separate components can be traced to maximum radii $\sim$4--10 kpc.  The ionized outflowing gas often extends as far as the detected stellar continuum in the host, indicating that outflows can expel gas at least to the outer regions of the stellar disks, if not beyond.  Earlier studies using integral field spectroscopy with smaller fields of view often found that the outflows filled the FOV, and as such the radial extents were lower limits.  Here we show that ionized outflows in quasars have extents up to $\sim$10 kpc; the median radial extent in our sample of 13 outflows is 7 kpc. Deep data cubes with the latest generation of wide-field IFS are powerful tools for tracing the full extent of outflowing gas in quasars \citep{2025ApJ...992...42R}.

The size of the extended and highly-ionized narrow-line region (NLR) in AGN, as delineated by \oiii, is known to correlate weakly with the luminosity of the NLR. It may correlate more strongly with the size of the NLR when considering only outflowing gas \citep{Sun17,Kang18}. In Figure~\ref{ro3-vs-lo3}, we compare the measured radial sizes of the broad and/or blueshifted \oiii\ regions in our sample with those from other samples of nearby AGN \citep{2013MNRAS.430.2327L,Harrison14,2014MNRAS.442.1303L,Bae17,Sun17,ros18,Fischer18,2018MNRAS.478.2438S}. We also compare to determinations of the $R_\mathrm{NLR}$ vs. $L_\mathrm{[OIII]}$ relation from several sources \citep{2011ApJ...732....9G,2014MNRAS.443..755H,Fischer18}, where $L_\mathrm{[OIII]}$ is the observed luminosity, without correction for extinction.  The differences among these determinations reflect different sample selections and sensitivity to extended \oiii\ emission. For the AGN in our sample with multiple outflows detected, we show the radial extent of the largest outflow.

The literature data show significant dispersion below the uppermost \oiii\ radius-luminosity relation of \citet{2011ApJ...732....9G}. The lowest radial measurements arise either from measurements with the {\it Hubble Space Telescope}, which is not optimized to low surface-brightness, extended emission; or from measurements using long-slit spectra, which cannot capture the full 2D projected distribution of ionized gas. The highest radii come from measurements with integral-field spectrographs.

Our new KCWI data increase the previously measured \oiii\ outflow radius and total luminosity in specific AGN by factors of 2--10 and are the highest measured values at fixed luminosities. The exception is I~Zw~1, for which similarly deep MUSE measurements exist \citep{2022ApJ...935...72M}. For this source, the MUSE FOV is greater than that of KCWI and the NLR is detected to larger radius with MUSE. Our measured radii are consistent with the NLR relation of \citet{2011ApJ...732....9G}, which was derived based solely on size and luminosity, not kinematics.

The radial extents measured in this paper should be considered lower limits, as with higher $S/N$ data, lower surface brightness outflows might be detected at larger radius.  Additionally, emission lines are sensitive to the density of the gas squared, such that lower density gas, if it is diffuse and more extended, may not be detected in emission.  Such gas could be detected at larger radius in absorption against a background source, if available. 

The ionized outflows detected in this sample have a wide range of physical properties, listed in Table 4.  The outflow in I Zw 1---the only narrow-line Type 1 AGN in our sample---in particular has properties that are orders of magnitude higher than in the other AGN, all of which are Type 2.  Across the full sample of seven quasars, the median values in each quantity are: mass, $4.1\times 10^7$\,\msun; mass outflow rate, 4\,\msunyr; kinetic energy, $2.6 \times10^{56}$\,erg; energy outflow rate, $1.9 \times10^{41}$ erg\,s$^{-1}$; and momentum flux, $2.4 \times10^{34}$\,dyn. These values are consistent with other estimates in nearby quasars, assuming comparable electron densities \citep[e.g.,][]{Rupke17,Harrison14,Bianchin26}.

\begin{figure*}[t!]
    \centering
    \includegraphics[width=0.85\textwidth]{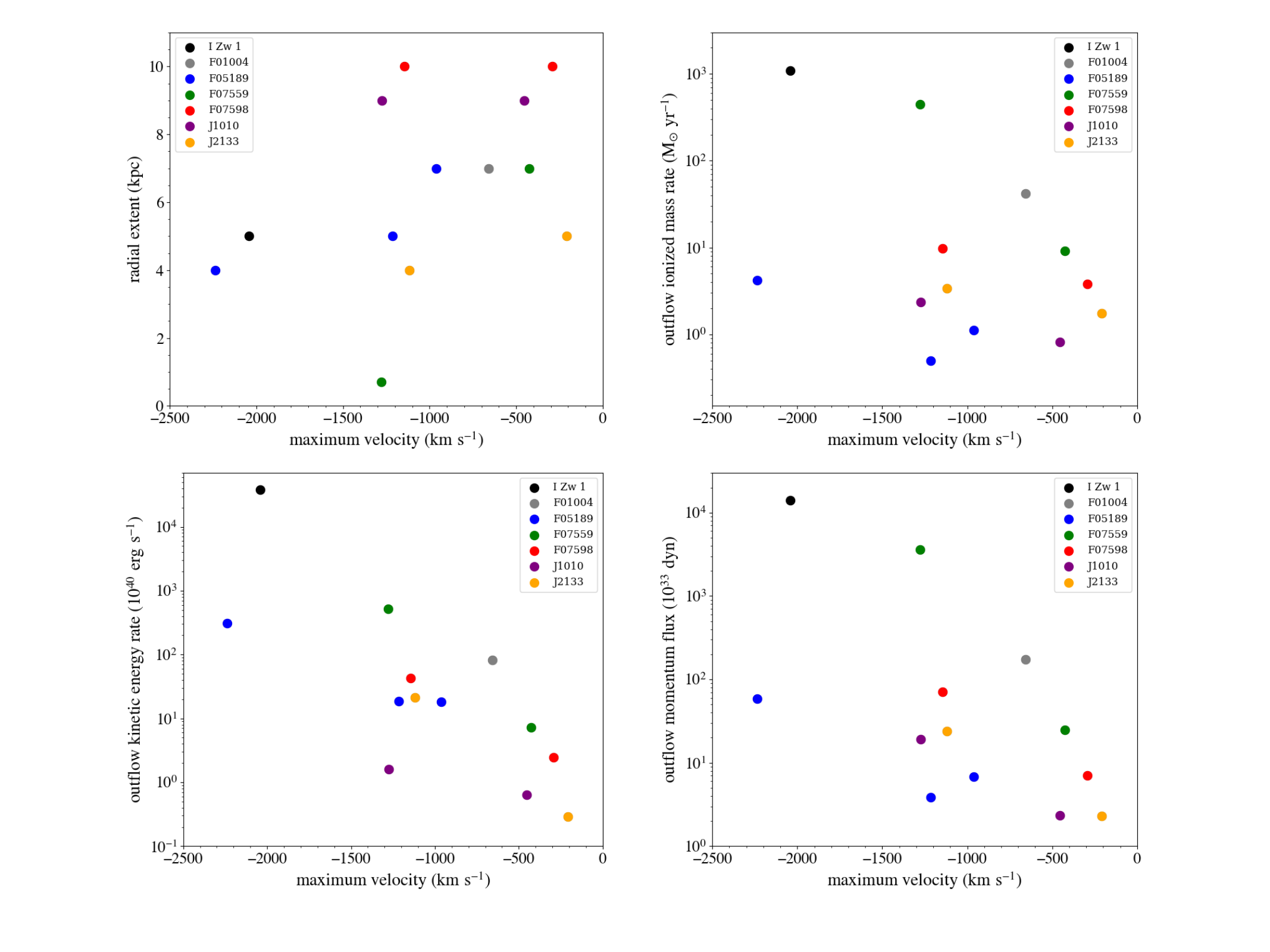}
    \caption{Trends between outflow properties derived for our quasar sample, shown as a function of the maximum outflow velocity for each outflow component in each source. Shown are the radial extent of each outflow component (upper left), estimated ionized mass outflow rate (upper right); estimated outflow kinetic energy rate (lower left); estimated outflow momentum flux (lower right).  As both I Zw 1 and F01004 have only a single outflow component detected, they are shown with one point each; the rest of the sources have multiple points corresponding to each outflow component.}
    \label{fig:outflow_props}
\end{figure*}

\subsection{Incidence and Properties of Multiple Outflows within the Same Galaxy}

Multiple outflowing components are detected within the same galaxy in five of the seven (71$\pm$17\%) quasars in this study. If we interpret these as separate outflow episodes, this suggests it is common for quasars to have multiple ionized outflows on $>$1 kpc scales. While most of the individual outflows detected here correlate with outflows found in previous studies of the same galaxies, previous studies identified only a single outflow component in J2133 \citep{Kang18}.  In models of radiation-driven black hole winds originating at the accretion disk, gas flows radially outward as a direct consequence of high accretion rates onto the central black hole \citep{1995ApJ...451..498M,2000ApJ...543..686P,2003MNRAS.345..657K}. If these small-scale winds in turn drive the motion of \oiii-emitting clouds at larger radius \citep{2015ARA&A..53..115K}, the multiple outflows can be traced to periods of high accretion separated in time. This variable accretion would also appear as AGN variability.

In this picture, we can estimate a rough timescale for each outflow from the maximum velocities and radial extents, $\sim$\vmax/R, assuming a constant velocity with time. We find that the median inferred timescale is 7$\times10^6$ years. These outflows have likely not traveled at constant velocity to kpc distances, but this timescale is consistent with the distinct outflows being driven by episodic AGN activity with shorter duration \citep{2011ApJ...737...26N,2015MNRAS.451.2517S}.

It is clear from the spectra presented here that high $S/N$ data reaching low surface brightness are needed to detect multiple outflows in a given source, as the faster outflows often have larger \vsig \ values and lower peak flux; these faster outflows would not be detected in low $S/N$ data. Additionally, IFS data yield spatial maps of the kinematics in each outflow, which constrains their time evolution.

Figure \ref{fig:outflow_props} shows the properties of each outflow detected here as a function of the maximum outflow velocity, for the outflow radial extent (upper left), ionized mass outflow rate (upper right), kinetic energy outflow rate (lower left), and momentum flux (lower right).  In the upper left panel we see that when there are multiple outflows detected within the same galaxy, the faster outflow has either the same or smaller radial extent than the slower outflow.  While being mindful of the small sample size shown here, these results are consistent with the idea that the faster outflows could have been launched more recently and have not has as much time to travel as far as the slower, potentially older, outflows.  This is consistent with AGN-driven outflows decelerating as they reach $>$kpc scales within their host galaxies.

The other panels in Figure \ref{fig:outflow_props} shows trends of increasing outflow mass rate, kinetic energy rate, and momentum flux with an increase in the maximum outflow velocity.  While not shown here, a similar trend is seen with outflow kinetic energy. The only derived parameter for which we do not find a trend with the maximum outflow velocity is the ionized mass in the outflow.

 For those AGN with multiple outflows, the faster outflows have higher mass outflow rates, kinetic energy rates, and momentum fluxes. While the faster outflows do not necessarily have more ionized mass, their higher velocities lead to larger mass outflow rates, energy and energy outflow rates, and momentum fluxes. To quantify this trend, we compare the physical properties of the faster outflows to those of the slower outflows {\it within a given galaxy} by taking the ratio of, for example, the mass outflow rate of the c3 component in J2133 to that of the c2 component in the same galaxy.  Doing this, we find that the median of the ratio of the mass outflow rate in the faster to slower outflows within a given AGN is 3, while the median of the ratio of the kinetic energy in the faster to slower outflows is 8, the median of the ratio of the outflow energy rate is 39, and the median of the ratio of the momentum flux is 10.  The faster outflows within a given AGN therefore have 3 times greater mass outflow rates, 10 times more energy, 18 times higher energy outflow rates, and 10 times more momentum flux, on average, than the slower outflows within the same galaxy.

\section{Summary and Conclusions}
\label{sec:concl}

In this study we present new Keck/KCWI IFS data of ionized gas outflows in 7 low redshift ($z\lesssim0.1$) quasars, 6 of which are Type 2 and 1 of which is a NL Type 1.  By simultaneously fitting the stellar emission and continuum light from the quasar  as well as line emission from multiple strong lines we are able to map the extent, morphology, and kinematics of outflowing gas in these galaxies.  We perform emission line fits to the \oii, \hg, \hb, and \oiii \ emission simultaneously, allowing for up to four distinct kinematic components, one of which traces gas in the host galaxy.  We present spatial maps of the \oiii/\oii \ line emission ratios for each kinematic component as well as maps of the velocity centroid and width (\vsig) in each spaxel for ionized gas in each component.

The main conclusions from our analysis are as follows:
\begin{enumerate}
\item Ionized gas outflows in low redshift quasars have radial extents of $\sim4-10$ kpc, often reaching extents that are comparable to the stars in the host galaxy.  The radial extents are consistent with trends observed between the NLR extents of quasars as a function of bolometric luminosity, indicating that the outflows reach comparable radii as the NL emission.

\item Five of the seven quasars (71\%) in this sample display multiple, distinct ionized outflows on these kpc scales with substantially different kinematic properties.  It is common for the outflows to reach velocities $>$1000 \kms \ with \vsig \ values of up to $\sim$600 \kms.  

\item For the quasars with multiple outflows, the faster outflows have comparable or somewhat smaller radial extents, consistent with the faster outflows potentially being younger. 

\item Within this sample there are trends for the ionized mass outflow rates, kinetic energy outflow rates, and momentum fluxes of the faster outflows to be substantially higher than the values for the slower ionized outflows.

\end{enumerate}

This work highlights the power of sensitive IFS observations over a wide FOV of AGN-driven outflows to both trace ionized outflows to large scales and to separate the kinematics of gas in the host galaxy from outflowing gas.  Having observations of and emission line fits to the \oii \ and \oiii \ lines in particular allows for robust identification and separation of emission from the host galaxy (which dominates the \oii \ emission) and the AGN photoionized outflowing gas (which is traced by \oiii).  This study further showcases how IFS data can be used to spatially map the extent and kinematics of distinct ionized outflows that exists on kpc scales within the same AGN.  Further IFS observations on larger samples of luminous AGN, both at low and higher redshift, would be useful to develop a statistical understanding of the properties of ionized gas outflows from AGN-driven feedback, which is needed to reveal the impact that these outflows have on their host galaxies.

\section*{Acknowledgements}

A.~L.~C. acknowledges support from the Ingrid and Joseph W. Hibben endowed chair at UC San Diego. The data presented herein were obtained at the W. M. Keck Observatory, which is operated as a scientific partnership among the California Institute of Technology, the University of California, and the National Aeronautics and Space Administration. The Observatory was made possible by the generous financial support of the W. M. Keck Foundation. The authors wish to recognize and acknowledge the very significant cultural role and reverence that the summit of Maunakea has always had within the indigenous Hawaiian community. We are most fortunate to have the opportunity to conduct observations from this mountain.

\bibliography{references}{}
\bibliographystyle{aasjournal}

\end{document}

%% file: targets_table.tex
\begin{deluxetable*}{ccrccccc}
  \tablecaption{Targets and Observations\label{tab:targets}}

  \tablehead{\colhead{Galaxy} & \colhead{RA} & \colhead{Dec} & \colhead{$z$} & \colhead{log\,($L_\mathrm{[OIII]}$/$\mathrm{erg} \ \mathrm{s}^{-1}$)} & \colhead{log\,($L_\mathrm{bol}/L_{\odot}$)} & \colhead{Ref.} & \colhead{Exp. Time (s)} }

  \startdata
    I~Zw~1        & 00:53:34.90 & $+$12:41:36 & 0.0604 & \nodata & 12.07   & 1 & 2550 \\[0.05cm]
    F01004$-$2237 & 01:02:50.00 & $-$22:22:00 & 0.1179 & 41.40   & 12.36   & 2 & 2880 \\[0.05cm]
    F05189$-$2524 & 05:21:01.40 & $-$25:21:45 & 0.0424 & \nodata & 12.22   & 1 & 3240 \\[0.05cm]
    F07559$+$5058 & 07:59:40.90 & $+$50:50:24 & 0.0544 & 42.35   & \nodata & 3 & 2500 \\[0.05cm]
    F07598$+$2601 & 08:02:52.93 & $+$25:52:56 & 0.0809 & 42.35   & \nodata & 3 & 2520 \\[0.05cm]
    J1010$+$0612  & 10:10:43.36 & $+$06:12:01 & 0.0983 & 42.30   & 12.02   & 4 & 2520 \\[0.05cm]
    J2133$+$0712  & 21:33:33.32 & $-$07:12:49 & 0.0867 & 41.30   & 11.82   & 5 & 3840 \\[0.05cm]
   \enddata
    \tablecomments{[\ion{O}{3}] 5007~\AA\ luminosities are from the sample outflow references, where given. Bolometric luminosities are independent measurements based on IR data \citep{2009ApJS..182..628V, Harrison14,Sun17}. References: (1) \citet{Rupke17}; (2) \citet{2018MNRAS.478.2438S};(3) \citet{Fischer18}; (4) \citet{Harrison14}; (5) \citet{Sun17}.}
  
\end{deluxetable*}

%% file: outflow_properties_table.tex
\begin{deluxetable*}{lccrrrr}
\tablecaption{Observed Properties by Kinematic Component \label{tab:kinematics}}
\tablehead{
\colhead{Target} & 
\colhead{$N_{\rm comp}$} & 
\colhead{Component} & 
\colhead{$v_{50,\rm med}$} & 
\colhead{$\sigma_{\rm med}$} & 
\colhead{$v_{\max,\rm med}$} & 
\colhead{$r_{\max}$} \\
 & & & \colhead{(km s$^{-1}$)} & \colhead{(km s$^{-1}$)} & \colhead{(km s$^{-1}$)} & \colhead{(kpc)}
}
\startdata
I Zw 1 & 2 & 1 & 0  & 50  & -171   & $>$9 \\
       &   & 2 & -543 & 732 & -2043 & 5 \\[0.15cm]
F01004-2237 & 2 & 1 & 0 & 87  & -200  & $>$18 \\
                &   & 2 & -94 & 330 & -658 & 7 \\[0.15cm]
F05189-2524 & 4 & 1 & 0  & 118 & -242    & 6 \\
            &   & 2 & -540  & 207 & -961  & 7 \\
            &   & 3 & -962 & 126 & -1214  & 5 \\
            &   & 4 & -1043 & 646 & -2238 & 4 \\[0.15cm]
F07559+5058 & 3 & 1 & 0  & 64  & -148   & 9 \\
            &   & 2 & 59  & 262 & -426   & 7 \\
            &   & 3 & 29 & 654 & -1279   & $<$3 \\[0.15cm]
F07598+2601 & 3 & 1 & 0  & 51  & -134   & 13 \\
            &   & 2 & 69  & 186 & -292  & 10 \\
            &   & 3 & -122 & 479 & -1145 & 10 \\[0.15cm]
J1010+0612 & 3 & 1 & 0  & 60  & -170   & 10 \\
           &   & 2 & -55  & 194 & -454  & 9 \\
           &   & 3 & -17 & 630 & -1274  & 9 \\[0.15cm]
J2133+0712 & 3 & 1 & 0  & 50  & -114  & 9 \\
           &   & 2 & 25  & 114  & -208 & 5 \\
           &   & 3 & -181 & 471 & -1118 & 4 \\[0.05cm]
\enddata
    \tablecomments{For each kinematic component the velocity centroid (\vcen), width (\vsig), maximum velocity (\vmax), and radial extent ($r_{max)}$ are listed.  The radial extent has been corrected for the seeing.}
\end{deluxetable*}

%% file: oiii_luminosities.tex
\begin{deluxetable}{lccc}
\tablecaption{\oiii \ 5007 \AA \ Luminosities \label{tab:oiii_lum}}
\tablehead{
\colhead{Target} & 
\colhead{Component} & 
\colhead{$L_{\oiii}^{obs}$} & 
\colhead{$L_{\oiii}^{corr}$} \\
\colhead{} &
\colhead{} &
\colhead{$10^{42}$} &
\colhead{$10^{42}$} \\
\colhead{} &
\colhead{} &
\colhead{erg s$^{-1}$} &
\colhead{erg s$^{-1}$}
}
\startdata
I Zw 1 & total & 1.57 & 46.5 \\
       & 1     & 0.07 & 0.21 \\
       & 2     & 1.55 & 46.1 \\[0.15cm]
F05189-2524 & total & 0.60 & 2.44 \\
       & 1     & 0.03 & 0.12 \\
       & 2     & 0.29 & 1.20 \\
       & 3     & 0.06 & 0.25 \\
       & 4     & 0.19 & 0.79 \\[0.15cm]
F07559+5058 & total & 3.72 & 82.8 \\
       & 1     & 0.55 & 12.3 \\
       & 2     & 1.08 & 24.0 \\
       & 3     & 2.07 & 46.1 \\[0.15cm]
F07598+2601 & total & 4.61 & 39.3 \\
       & 1     & 1.19 & 10.1 \\
       & 2     & 1.76 & 15.0 \\
       & 3     & 1.64 & 14.0 \\[0.15cm]
J1010+0612  & total & 3.00 & 3.00 \\
       & 1     & 0.09 & 0.09 \\
       & 2     & 0.89 & 0.89 \\
       & 3     & 1.81 & 1.81 \\[0.15cm]
J2133+0712  & total & 0.42 & 7.06 \\
       & 1     & 0.09 & 1.46 \\
       & 2     & 0.14 & 2.29 \\
       & 3     & 0.19 & 3.20 \\[0.05cm]
\enddata
    \tablecomments{[\ion{O}{3}] 5007 \AA \ luminosities are provided both for the observed values and the dust-corrected values, using the $E(B-V)$ measurements derived from the Balmer decrements presented in Section 4.}
\end{deluxetable}

%% file: derived_properties_table.tex
\begin{deluxetable*}{lcccrrrrrr}
\tablecaption{Derived Ionized Gas Outflow Properties \label{tab:energetics}}
\tablehead{
\colhead{Target} & 
\colhead{Component} & 
\colhead{$E(B-V)$} &
\colhead{$L_{\mathrm{H}\beta}^{\mathrm{obs}}$} &
\colhead{$L_{\mathrm{H}\beta}^{\mathrm{corr}}$} & 
\colhead{$M_{\rm ion}$} & 
\colhead{$\dot{M}$} & 
\colhead{$E$} & 
\colhead{$\dot{E}$} & 
\colhead{$\dot{P}$} \\
\colhead{} &
\colhead{} &
\colhead{} &
\colhead{$10^{40}$} &
\colhead{$10^{40}$} &
\colhead{$10^{6}$} &
\colhead{} &
\colhead{$10^{55}$} &
\colhead{$10^{40}$} &
\colhead{$10^{33}$} \\
\colhead{} &
\colhead{} &
\colhead{} &
\colhead{erg s$^{-1}$} &
\colhead{erg s$^{-1}$} &
\colhead{$M_\odot$} &
\colhead{$M_\odot$ yr$^{-1}$} &
\colhead{erg} &
\colhead{erg s$^{-1}$} &
\colhead{dyn}
}
\startdata
I Zw 1      & 2 & 1.06 $\pm$0.26 & 92.2 & 2,767 & 2,629 & 1098 & 10,913 & 38,410 & 14,147 \\[0.10cm]
F01004-2237 & 2 & 0.98 $\pm$0.18 & 20.1 & 460   & 437   & 42.0 & 188   & 81.9    & 174 \\[0.10cm]
F05189-2524 & 2 & 0.44 $\pm$0.17 & 2.08 & 8.38  & 8.0   & 1.12 & 7.31  & 18.3   & 6.77 \\
            & 3 &      & 0.53 & 2.13  & 2.0   & 0.50 & 2.97  & 18.5   & 3.85 \\
            & 4 &      & 1.91 & 7.70  & 7.3   & 4.18 & 36.4  & 308    & 59.1 \\[0.10cm]
F07559+5058 & 2 & 0.97 $\pm$0.40 & 7.05 & 156   & 148   & 9.20 & 26.7  & 7.29   & 24.7 \\
            & 3 &      & 11.3 & 250   & 238   & 444  & 387   & 519  & 3581  \\[0.10cm]
F07598+2601 & 2 & 0.67 $\pm$0.23 & 15.7 & 135   & 128   & 3.83 & 10.9  & 2.44   & 7.06 \\
            & 3 &      & 10.2 & 87.9  & 83.5  & 9.78 & 109   & 43.1    & 70.6 \\[0.10cm]
J1010+0612  & 2 & 0.00 & 16.5 & 16.5  & 15.7  & 0.81 & 3.22  & 0.64   & 2.32 \\
            & 3 &      & 17.2 & 17.2  & 16.4  & 2.37 & 26.4  & 1.62    & 19.0 \\[0.10cm]
J2133+0712  & 2 & 0.88 $\pm$0.32 & 2.56 & 43.3  & 41.2  & 1.75 & 1.77  & 0.29   & 2.30 \\
            & 3 &      & 0.74 & 12.4  & 11.8  & 3.37 & 14.7  & 21.5    & 23.8 \\
\enddata
\end{deluxetable*}